\documentclass[aps,prb,reprint,amsfonts,showpacs,letterpaper,floatfix]{revtex4-1}
\pdfoutput=1
\usepackage{CJK}
\usepackage{graphicx, multirow,bm, longtable,subfigure}
\usepackage{hyperref}
\usepackage{fonttable}
\usepackage{amsmath}
\usepackage{amsfonts}
\usepackage{color}
\begin{document}
\begin{CJK*}{UTF8}{bsmi}  
	\title{Compressive sensing as a new paradigm for model building}
	\author{Lance J. Nelson}
	\affiliation{Department of Physics and Astronomy, Brigham Young University, Provo, Utah
          84602, USA} 
	\author{Fei Zhou (周非)}
	\affiliation{Department of Materials Science and Engineering, University of California,
          Los Angeles, California 90095, USA}  
	\author{Gus L. W. Hart} 
	\affiliation{Department of Physics and Astronomy, Brigham Young University, Provo, Utah
          84602, USA} 
	\author{Vidvuds Ozoli\c{n}\v{s}} 
	\affiliation{Department of Materials Science and Engineering, University of California,
          Los Angeles, California 90095, USA}
        \email{vidvuds@ucla.edu} 
	\date{\today} 
\begin{abstract}
  The widely-accepted intuition that the important properties of
  solids are determined by a few key variables underpins many methods
  in physics. Though this reductionist paradigm is applicable in many
  physical problems, its utility can be limited because the intuition
  for identifying the key variables often does not exist or is
  difficult to develop. Machine learning algorithms (genetic
  programming, neural networks, Bayesian methods, etc.) attempt to
  eliminate the { \it a priori} need for such intuition but often do
  so with increased computational burden and human time. A
  recently-developed technique in the field of signal processing,
  compressive sensing (CS), provides a simple, general, and efficient
  way of finding the key descriptive variables.  CS is a new paradigm
  for model building---we show that its models are more physical
  and predict more accurately than current state-of-the-art approaches, 
  and can be constructed at a fraction of the computational cost and user effort.

\end{abstract}

\maketitle
\end{CJK*}

\section{Inroduction}
\label{sec:intro}

Physical intuition and experience suggest that many important
properties of materials are primarily determined by just a few key
variables. 
For instance, the crystal structures of intermetallic compounds have
been successfully classified into groups (so-called structure maps)
according to the properties of the constituent
atoms.\cite{Zunger1980,Villars1983,Pettifor1984,Pettifor1986} The
widely known Miedema rules relate alloy formation energies to atomic
charge densities and electronegativities.\cite{Miedema1975} Most
magnets can be described using a Heisenberg model with only a few
short-ranged exchange interactions,\cite{Kormann2008}
and the formation energies of multicomponent alloys can be efficiently
parameterized using generalized Ising models (cluster expansions) with
a finite number of pair and multibody
interactions.\cite{sanchez1984generalized,fontaine1994cluster,Zunger1994}
In all these cases, enormous gains in efficiency and conceptual
clarity are achieved by building models which express the quantity of
interest (typically, total energy) in a simple, easy-to-evaluate
functional form. These models can then be used to perform realistic
simulations at finite temperatures, on large systems, and/or over long
time scales, significantly extending the reach of current
state-of-the-art quantum mechanics based methods.

The conventional approach to model building starts by selecting a
small, physically motivated basis set which describes the
configuration of the system. The target properties are then expressed
in terms of these basis functions and the unknown coefficients are
determined by performing least-squares fits to the calculated or
experimentally measured data. While conceptually simple, this method
is often difficult to use in practice. First, the number of unknown
coefficients has to be smaller than the number of data points, which
precludes the use of very large basis sets. Second, least-squares
fitting is susceptible to noise, and there is often the possibility of
``over-fitting''---the model is trained to reproduce the fitting data,
but performs poorly in a predictive capacity. Finally, finding the the
optimal finite basis set is an $NP$-hard problem, i.e., the solution
time increases faster than polynomial with the number of possible
basis functions. To keep the number of coefficients smaller than the
amount of data, one must choose, based on physical intuition, which
basis functions to keep.  This physical intuition in many cases may be
unavailable and/or difficult to develop; hence there is no clear
path to achieve systematic improvement. Recent years have seen numerous
attempts to use machine learning algorithms (genetic programming,
neural networks, Bayesian methods, etc.) to decrease the role of
intuition in
model-building.\cite{Fischer:2006ip,PSSB:PSSB200945246,1749-4699-2-1-015006,stefano_scintillator,Woodley:2008hha,jansen2008bayesian,mueller2009bayesian,cockayne2010building}

We show that a recently developed technique in the field of signal
processing, compressive sensing (CS),\cite{candes2008introduction}
provides a simple, general, and efficient approach to 
model-building.\cite{alquraishi2011direct} Instead of attempting to develop
physical intuition for which coefficients will be most relevant, the
CS framework allows the inclusion of essentially all possible basis
functions. Using very large basis sets eliminates the need to use
physical intuition to construct smaller ones. Furthermore, CS is
computationally efficient for very large problems, robust even
for very noisy data, and its models predict more accurately than
current state-of-the-art approaches.


\section{Compressive sensing: an illustration}
\label{sec:Fourier}

\begin{figure*}
\noindent
\parbox{0.70\linewidth}{
\includegraphics[width=.49\linewidth]{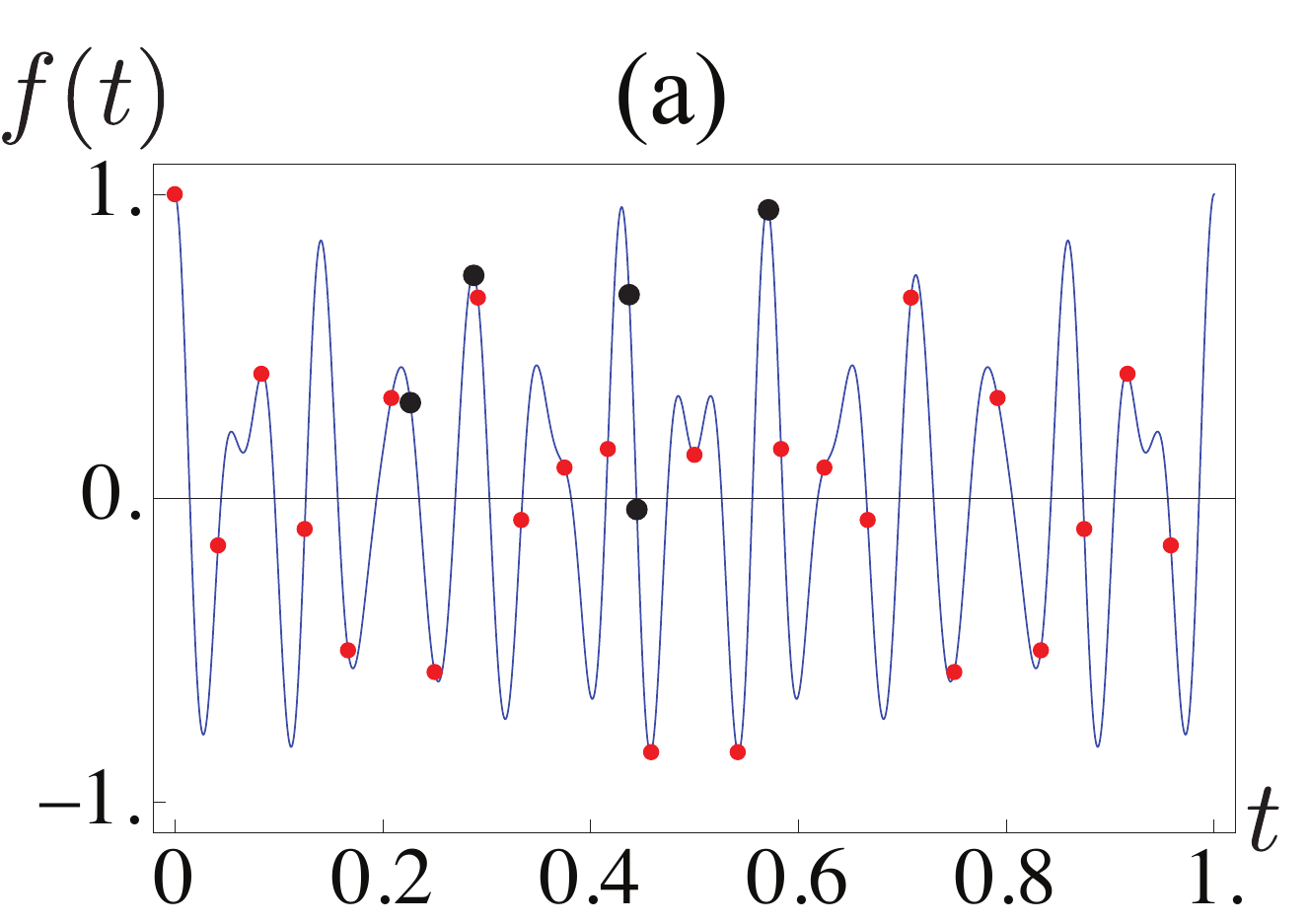}\hfill\includegraphics[width=.50\linewidth]{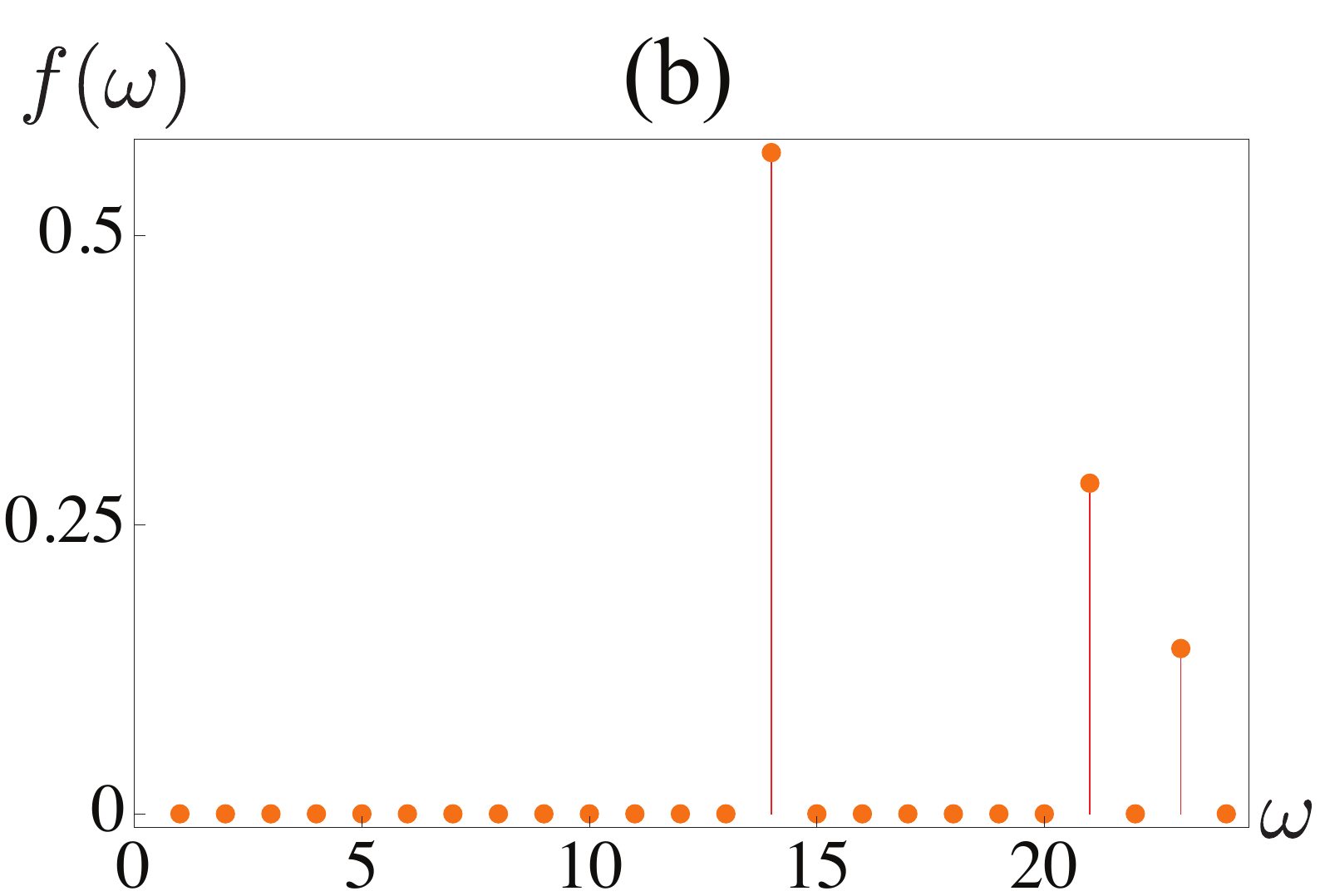}}
\hfill
\parbox{0.27\linewidth}{
  \caption{    \label{fig:time_signal}
    (a) A sparse signal (blue line) like that of Eq.~\ref{eq:time_signal}, uniform samples
    of the signal at the Nyquist frequency (red dots), and a few random samples (black circles). The
    signal is composed of only 3 non-zero frequencies.  
    (b) Exact recovery of the frequency components of the signal using compressive sensing.
    	}
    }
\end{figure*}

Before demonstrating the power of compressive sensing for building physical models, we first illustrate the concept itself with a simple time series. Discussion of compressive sensing requires the definition of $\ell_{p}$ norms:
\begin{equation}
\label{eq:p-norm}
\lVert u \rVert_{p} = 
\left( \sum_i |u_i |^{p}\right)^{1/p},
\end{equation}
of which the $\ell_{1}$ (taxicab or Manhattan distance) and $\ell_{2}$
(Euclidean; subscript 2 often omitted) norms are special cases. The
number of non-zero elements of $\vec{u}$ is often (improperly)
referred to as the $\ell_{0}$ ``norm'' even though it is not a norm in a
strict mathematical sense.

In the signal processing community, compressive sensing is used
to recover sparse signals {\it exactly} with far fewer samples
than required by standard spectral techniques, such as the well-known
Fourier and Laplace transforms. Consider a signal like that shown in
Fig.~\ref{fig:time_signal}(a) which has the functional form:
\begin{equation}
\label{eq:time_signal}
f(t) = \sum_{n=1}^N u_n e^{i2\pi n t},
\end{equation}
where most of the coefficients, $u_n$, are zero (i.e., the signal is sparse). 
The Fourier transform
is mathematically equivalent to solving the matrix equation
\begin{equation} \label{eq:AuEQf}
\mathbb{A}\vec u = \vec f,
\end{equation}
where the matrix $\mathbb{A}$ is formed by the values of the Fourier
basis functions at the sampling times $t_{m}$, i.e., it consists of
rows of $n$ terms of the form $A_{mn}=e^{i2\pi n t_m}$, and $f_m
\equiv f(t_{m})$ is the sampled signal. The solution vector $\vec u$
contains the relative amounts of the different Fourier components, as
shown in Fig.~\ref{fig:time_signal}(b). Capturing all relevant
frequency components of the signal using Fourier transform techniques
requires the signal to be sampled regularly and at a frequency at
least as high as the Nyquist frequency [shown as red points in
Fig.~\ref{fig:time_signal}(a)], a severe restriction stemming from the
requirement that the linear system Eq. (\ref{eq:AuEQf}) should not be
underdetermined.

However, the main idea of compressive sensing is that,when the signal
is sparse, one should be able to recover the exact signal with a
number of measurements that is proportional to the number of nonzero
components, i.e., with far fewer samples than given by the Nyquist
frequency.  Conceptually, this could be done by searching for a
solution that reproduces the measured time signal exactly {\it and}
has the minimum number of non-zero Fourier components. Unfortunately,
this formulation results in a discrete optimization problem, which
cannot be solved in polynomial time. Compressive sensing recasts the
problem as a simple minimization of the $\ell_1$ norm of the solution,
subject to the constraint given by Eq.~(\ref{eq:AuEQf}) above:

\begin{equation}
\label{eq:const_min}
\min_u\{\lVert \vec u\rVert_1:\mathbb{A}\vec u=\vec f\},
\end{equation}
where $\lVert \vec u \lVert_1 = \sum_i | u_i | $ is the $\ell_1$-norm defined in Eq.~(\ref{eq:p-norm}). In
other words, one seeks to minimize the sum of the components of the
solution vector $\vec u$ subject to the condition that the measured
signal is reproduced exactly; this constitutes the so-called basis
pursuit problem. Eq.~(\ref{eq:const_min}) is a convex optimization
problem which can be solved efficiently (see Sec.~\ref{sec:L1}). 
We note here that optimization of the common
sum-of-squares ($\ell_{2}$) norm of $\vec u$ would result in a 
dense solution which may deviate considerably from the original
signal.\cite{candes2008introduction}

As a simple illustration, the exact decomposition of an example
function, shown in Fig.~\ref{fig:time_signal}, was possible via
compressive sensing with only 5 random samples (black dots in figure
\ref{fig:time_signal}) of the signal, instead of the 24 equally-spaced
samples (red dots in figure \ref{fig:time_signal}) needed for a
discrete Fourier transform. Quite generally, a mathematical theorem
proven by Candes, Romberg, and Tao\cite{candes2006robust} guarantees
that, with an overwhelming probability, any sparse signal with $S$
nonzero components can be recovered from $M \sim S\log N$ random
measurements, where $N$ is the total number of sensing basis
functions.  This very powerful result is the mathematical foundation
of compressive sensing. 

Another practically important feature of compressive sensing is the
ability to tolerate noise in the input data and to deal with signals
that are only approximately sparse, i.e., are dominated by a few large
terms, but also contain a large number of smaller contributions; this
is the case in almost all physics applications.  It has been proven
that, if the sensing matrix $\mathbb{A}$ obeys the so-called {\it
  restricted isometry property\/} (RIP), an accurate reconstruction of
the signal from highly under-sampled measurements can be achieved also
in the presence of both random and systematic
noise.\cite{candes2008introduction,candes2006robust} The RIP criterion
is automatically satisfied if the measurements are chosen randomly.
(see Sec.~\ref{sec:structures} for a detailed discussion)

When applying compressive sensing to model building, two tasks must be
accomplished: (i) a basis must be chosen, and (ii) the coefficients
associated with each basis function must be
determined. Mathematically, the problem is analogous to the simple
Fourier example considered above, with the sensing matrix $\mathbb{A}$
being determined by the values of the basis functions at the chosen
measurement points.  Below we illustrate the use of compressive
sensing on two cluster expansion (CE) models of configurational
energetics:\cite{sanchez1984generalized} (i) Ag-Pt alloys on a
face-centered cubic (fcc) lattice, and (ii) protein folding energies
in the so-called zinc finger motif. CE is chosen as an example because
it is conceptually simple, mathematically rigorous, and widely used in
the materials community to calculate temperature-composition phase
diagrams. Furthermore, CE is a stringent test case for compressive
sensing because a significant amount of effort has been expended
developing advanced model building techniques, which have been
implemented in sophisticated general-purpose computer
codes.\cite{vandeWalle2002,vandeWalle2009,lerch2009uncle,jansen2008bayesian,mueller2009bayesian,Johnson2004,cockayne2010building,Ceder2010}
 
\section{Cluster Expansion}
\label{sec:CE}

\subsection{Energy Model}

Since a formal mathematical description of CE can be found in the literature, here we only restate its main features and refer the reader to Refs.~\onlinecite{sanchez1984generalized,fontaine1994cluster,Zunger1994} for detailed explanations. The CE method uses a complete set of discrete basis functions, defined over clusters of lattice sites, which describe the occupation of each site and thus the entire atomic configuration on the crystal. The total energy is given by
\begin{equation}
\label{eq:CE}
E(\sigma) = E_0 + \sum_f \bar\Pi_f (\sigma) J_f,
\end{equation}
where $f$ represents symmetrically distinct clusters of lattice sites
(points, pairs, triplets, etc.), $\sigma$ denotes the atomic
configuration, usually expressed by a collection of pseudo-spin
variables $\{ S_i \}$ describing the type of atom at each lattice
site, and the cluster correlations $\bar\Pi_f (\sigma)$ are formed as
symmetrized averages of products of these pseudo-spin variables. The
key quantities in this approach are $J_f$, the effective cluster
interactions (ECI's): Given the ECI's, the energy of {\it any\/}
atomic configuration on the lattice can be calculated rapidly from
Eq.~(\ref{eq:CE}). Physical intuition based on the concept of
``near-sightedness''of screened interatomic interactions 
suggests that only clusters within a limited range and involving a
limited number of sites will have significant ECI's. The goal
then is to determine which of the clusters $f$, out of the myriads of
possible choices, contribute significantly to the total energy of the
system and to calculate the values of these coefficients. 

Currently, the most popular approaches are based on the so-called
structure inversion method (SIM),\cite{ConnollyWilliams} where a
limited number of quantum-mechanics-based total energy calculations
are used to determine $E(\sigma)$ on the left-hand side of
Eq.~(\ref{eq:CE}). The cluster interactions $J_f$ are truncated
according to some recipe and their values are determined by
least-squares fitting to the training set energies $E(\sigma)$. The accuracy of 
the resulting CE depends crucially on the chosen truncation method. 
Including too few interactions leads to poor predictive power because important
interactions are not accounted for (``under-fitting'', while choosing
too many parameters $J_f$ results in spurious interactions and an
associated decrease in predictive accuracy (``over-fitting''). Use of
the least-squares fitting necessarily requires that the number of
structures must exceed the number of candidate ECIs, which
is the CE analogue of the Nyquist frequency in signal
processing.

In modern practice, the trial ECI's are chosen by scanning over many
possible sets of clusters while attempting to minimize the predictive
error. Ideally, the predictive error should be calculated 
as the root mean square (RMS) deviation between the density functional theory (DFT) and CE-predicted
energies over a separate ``hold-out'' set of structures that are not used in fitting.
This approach would require tens or hundreds of additional DFT
calculations and is therefore seldom used in practice. Leave-one-out 
cross-validation (LOOCV) or $k$-fold cross-validation scores are commonly
used as proxies for the predictive error since they do not require the
construction of a separate hold-out set.\cite{vandeWalle2002}

Starting from an initial set of ECI's (e.g., empty, point, and
nearest-neighbor pair clusters), a typical procedure for improving the
model attempts to add and/or substitute clusters into the current set,
keeping changes if the predictive error is found to decrease. The
procedure is terminated when none of the attempted changes produce an
improvement in the predictive accuracy.  Unfortunately, there is no
guarantee that this process will truly result in the optimal set of
ECI's because it is practically impossible to solve the $NP$-hard
discrete optimization (DO) problem, especially if the number of
candidate ECI's is large, such as required for very accurate CE's or
in situations of low symmetry (e.g., near defects, surfaces,
nano-particles). As a result, with DO, one may miss important clusters
or add clusters that should not have been included in the model, which
may result non-physical ECI values. Additionally, adding/removing
clusters one-by-one is computationally expensive, requiring days to
finish when considering very large pools of candidate clusters.

Genetic algorithms have been used with some success, but they also
require large amounts of time to complete, especially for large
  cluster pools and fitting sets, and employ a host of tunable
  paramters.\cite{blum2005using,lerch2009uncle}
Design of efficient, numerically robust and physically accurate
methods for selecting the physically significant ECIs remains a
challenging problem.

Other researchers, in an attempt to avoid predictive
errors associated with incomplete discrete optimization, have sought to 
devise direct minimization methods  that automatically select ECIs only
if they are required to reproduce the energies of the training set. 
The first such approach was proposed by Laks, Wei, and
Zunger for pair interactions,\cite{DBLaks} who added a
distance-weighted $\ell_2$ norm of the pair interactions to the
objective function. However, this approach usually results in dense
sets of long-ranged pair ECI's and, more importantly, is
difficult to extend to multibody interactions.\cite{Bugaev2010,Besson2012,DiazOrtiz2008} 
Recently, a method based on
Bayesian statistics was introduced to automatically estimate
ECI's and shown to outperform several common methods
in low-symmetry situations.\cite{Mueller2009} However, it makes use 
of physical intuition to construct informative prior distributions, 
which are required for estimating the ECI values. 
It is desirable to develop methods that avoid the use of intuition 
since heuristic rules, derived from experience with a few 
specific systems, may not be universally valid. 

\subsection{Compressive sensing cluster expansion (CSCE)}
\label{sec:CSCE}


Here, we show that compressive sensing can be used to select the
important ECI's and determine their values {\it in one shot}. The
applicability of compressive sensing to CE is based on the
mathematical theorem of Candes, Romberg, and
Tao,\cite{candes2006robust} which guarantees that sparse ECI's can be
recovered from a limited number of DFT formation energies given
certain easy-to-satisfy properties of the matrix $\bar\Pi$ in
Eq.~(\ref{eq:CE}). Adopting the common assumption that the ``true'' physical 
ECI's are approximately sparse, this theorem guarantees 
that a good approximation will be found even 
in cases when the data has both random and systematic
noise, e.g., due to numerical errors in the DFT calculations or due to
interactions beyond the chosen energy resolution, see Sec.~\ref{sec:mu}.

There are two possible formulations for compressive sensing cluster expansion 
(CSCE), both of which enforce
the requirement that the cluster expansion should be as sparse as
possible, while resulting in a certain level of accuracy for the
training set. In the first approach, one may determine the optimal set
of ECI's from
\begin{equation}
\label{eq:CSCEv1}
J = \arg \min_J \left\{ || J ||_1 : || E - \bar\Pi J || \le \epsilon \right\},
\end{equation}
where $\ell_1$ norm of $J$'s is used as a proxy for the number of nonzero
ECI's. Solving the so-called 
LASSO problem Eq.~(\ref{eq:CSCEv1})\cite{Tibshirani1996,Chen1998} 
offers a mathematically strict way of constructing a minimal cluster 
set that reproduces the training set with a given accuracy.
Of course, over- (under-) fitting is still an
issue if $\epsilon$ is chosen too small (large), but it is physically
reasonable that, given the physical properties of the system and the
size of the training set, an optimal $\epsilon$ always exists.
Following common practice, optimal $\epsilon$ could be found 
either by minimizing the LOOCV score or the predictive RMS error over a
hold-out set.

Since the inequality constraint is inconvenient to enforce during
calculations,\cite{Tibshirani1996} here we follow common practice 
in signal processing and use an unconstrained approach which minimizes
the sum of an $\ell_1$ norm of the ECI's and a least-squares sum of
the fitting errors:
\begin{equation}
\label{eq:CSCEv2}
J = \arg \min_J  \mu || J ||_1 + \frac{1}{2} || E - \bar\Pi J ||^2,
\end{equation}
where $\mu$ is a parameter that controls the accuracy of the fit
versus the sparseness of the solution: higher values of $\mu$ will
result in sparser solutions and larger fitting errors (under-fitting), while very
small $\mu$ values will lead to dense solutions and degraded
predictive accuracy (over-fitting). It will be shown below in Sec.~\ref{sec:mu} 
that the optimal value of $\mu$ is proportional to the level of noise 
(random and systematic) in the calculated formation energies. 
Just like $\epsilon$ in Eq.~(\ref{eq:CSCEv1}), an optimal $\mu$ 
to avoid over- or under-fitting can
be chosen either by minimizing the LOOCV score or by minimizing the
rms prediction error for a separate hold-out set; it is shown below 
that both approaches result in very similar values of optimal $\mu$. 
Furthermore, in Sec.~\ref{sec:applications} we demonstrate 
that CSCE is not particularly sensitive to the precise value of 
$\mu$ and show that there is usually a range of  $\mu$'s 
that give ECI's of similar predictive accuracy. 

The main advantage of CSCE, Eqs.~(\ref{eq:CSCEv1}) \& (\ref{eq:CSCEv2}),
over current CE methods is that the $NP$-hard discrete optimization of
the truncated ECI set is replaced by convex optimization problems for
which exact solutions may be found in polynomial time. Furthermore, the
minimization of the $\ell_1$ norm of the solution also serves to
decrease the magnitude of the ECI's, leading to ``smoother'' ECI's,
increased numerical stability with respect to the noise in the
training data, and eventually more accurate predictions. In addition, the CSCE is
simple to implement and use, which will facilitate its widespread
adoption in solid state physics and other fields where configurational
energetics play a role. In Sec.~\ref{sec:applications} below, we 
illustrate the superior performance of CSCE using examples 
from bulk alloys (Ag-Pt) and biology (protein folding energetics).

\section{Practical aspects of $\ell_1$-based optimization}
\label{sec:L1}

In what follows, we review methods for solving the unconstrained minimization problem given by Eq.~(\ref{eq:CSCEv2}), which we rewrite as:
\begin{equation}
\label{eq:const_min_w_mu}
\min_u \mu \lVert \vec u \rVert_1+ {\textstyle \frac{1}{2}} \lVert \mathbb{A} \vec u - \vec f \rVert^2.
\end{equation}
Eq.~(\ref{eq:const_min_w_mu}) is referred to as the basis pursuit
denoising problem. It has a tunable parameter, $\mu$, which controls
the sparseness of the solution: smaller (larger) values of $\mu$
produce less (more) sparse solutions.

\subsection{Fixed-point continuation}
\label{sec:FPC}

The fixed-point continuation (FPC) method of Hale, Yin, and Zhang\cite{HaleYinZhang2007} is an iterative algorithm that starts from $\vec{u}^0=\mathbf{0}$ and attempts to improve the objective function by following the gradient of the $\ell_2$ term:
\begin{eqnarray}
\label{eq:FPC1}
\vec{g}^k = \mathbb{A}^T (\mathbb{A} \vec{u}^k - \vec{f})\\
\label{eq:FPC2}
u^{k+1}_n = \text{shrink} \left( u^k_n - \tau g^k_n , \mu \tau \right)
\end{eqnarray}
where $k=0,1,2,\ldots$ is the iteration number and the shrinkage operator is defined as
\begin{equation}
\text{shrink} \left( y, \alpha \right) := \text{sign} (y) \max \left( |y|-\alpha,0 \right).
\end{equation}
In other words, shrinkage decreases the absolute magnitude of $y$ by
$\alpha$ and sets $y$ to zero if $|y| \le \alpha$.  The iterations are
stopped when the $\ell_\infty$ norm, or maximum component value,
of the gradient drops below the shrinkage threshold, 
\begin{equation}
\label{eq:deltag}
\frac{1}{\mu} || \vec{g} ||_\infty - 1 < \delta_g, 
\end{equation}
and the change in the solution vector is sufficiently small, 
\begin{equation}
\label{eq:deltau}
\frac{\|| \vec{u}^{k+1} - \vec{u}^k ||}{|| \vec{u}^k ||} < \delta_u. 
\end{equation}
The sensing matrix should be normalized in such a way that the largest eigenvalue $\alpha_A$ of $\mathbb{A}^T \mathbb{A}$ is less than or equal to 1; this is easily accomplished by dividing both $\mathbb{A}$ and $\vec{f}$ by $\sqrt{\alpha_A}$. The step size $\tau$ in Eq.~(\ref{eq:FPC2}) is given by
\begin{equation}
\label{eq:tau}
\tau = \min (1.999,-1.665 \frac{M}{N} + 2.665),
\end{equation}
where $M$ and $N$ are the number of equations and the number of
expansion coefficients, respectively.

\subsection{Bregman iteration}
\label{sec:Bregman}

While FPC algorithm is generally applicable to any problem of type Eq.~(\ref{eq:const_min_w_mu}) and is guaranteed to converge, in practice it has a serious shortcoming: very small values of $\mu$ are needed to recover the exact solution to the basis pursuit problem without noise, Eq.~(\ref{eq:const_min}), which cause an associated increase in the number of FPC iterations. To alleviate the need to use small $\mu$'s, Yin {\it et al.\/}\cite{yin2008bregman} proposed an efficient iterative denoising algorithm for finding the solution to Eq.~(\ref{eq:const_min_w_mu}), which has the additional benefit of yielding the exact solution to the basis pursuit problem Eq.~(\ref{eq:const_min}) for zero noise. This so-called Bregman iteration involves the following two-step cycle:
\begin{eqnarray}
\label{eq:fk+1}
\vec{f}^{k+1} &=& \vec{f} + (\vec{f}^k - \mathbb{A} \vec{u}^k),\\
\label{eq:uk+1}
\vec{u}^{k+1} &=& \arg \min_u \mu \lVert \vec u \rVert_1+ 
				{\textstyle \frac{1}{2}} \lVert \mathbb{A} \vec u - \vec{f}^{k+1} \rVert^2,
\end{eqnarray}
starting from $\vec{f}^0=\mathbf{0}$ and $\vec{u}^0=\mathbf{0}$. A key
feature of the algorithm is that the residual after iteration $k$ is
added back to the residual vector $\vec{f}^{k+1}$ for the next
iteration, resulting in efficient denoising and rapid
convergence.\cite{yin2008bregman} Each minimization in
Eq.~(\ref{eq:uk+1}) can be performed using the fixed-point
continuation (FPC) method proposed by Hale, Yin, and
Zhang.\cite{HaleYinZhang2007} The main advantages of the Bregman
iteration are faster convergence and the ability to use $\mu$ 
values that are several orders of magnitude larger than those 
required for direct application of the FPC method.

\subsection{Split Bregman iteration}
\label{sec:SplitBregman}

For very large problems (i.e., large sensing matrices $\mathbb{A}$),
the FPC optimization steps in the Bregman iterative method progress
very slowly. The problem becomes severe when the
condition number computed from the nonzero eigenvalues of
$\mathbb{A}^T \mathbb{A}$ becomes large. Indeed, FPC is essentially a
steepest descent method combined with an $\ell_1$ shrinkage step, and
the number of required steepest descent iterations 
increases linearly with the ratio of the largest-to-smallest
eigenvalues of $\mathbb{A}^T \mathbb{A}$.\cite{Boyd} An improved Bregman
algorithm, which eliminates the hard-to-solve mixed $\ell_1$ and
$\ell_2$ minimization problem in Eq.~(\ref{eq:uk+1}), was proposed
by Goldstein and Osher.\cite{goldstein2009split} It carries
the name of ``split Bregman'' iteration because it splits off the
$\ell_1$ norm of the solution from the objective function and
replaces it with a new variable $\vec{d}$, which is designed to 
converge towards the $\ell_1$ term, 
$ \lim_{k \rightarrow \infty} ( \vec{d}^k - \mu \vec{u}^k ) = \mathbf{0}$. 
A new least-squares $\ell_2$ term is added to the objective function 
to ensure that $\vec{d} = \mu \vec{u}$ in the limit:
\begin{equation}
\label{eq:split_min}
\vec{u} = \arg \min_{u,d} \lVert \vec d \rVert_1+ {\textstyle \frac{1}{2}} \lVert \mathbb{A} \vec u - \vec f \rVert^2
+ {\textstyle \frac{\lambda}{2}} \lVert \vec d - \mu \vec u \rVert^2.
\end{equation}
A key advantage of this formulation is that the minimization involving the quadratic form $\lVert \mathbb{A} \vec u - \vec f \rVert^2$ does not contain $\ell_1$ terms and can be performed efficiently using standard convex optimization techniques, such as Gauss-Seidel or conjugate gradients (CG),\cite{Boyd} while the $\ell_1$ minimization with respect to $\vec{d}$ at a fixed $\vec{u}$ contains an $\ell_2$ term that is diagonal in the components of $\vec{d}$ and can be solved easily (see below). The full split Bregman iterative algorithm proceeds as follows:
\begin{eqnarray}
\label{eq2:uk+1}
\vec{u}^{k+1} &=& \arg \min_u {\textstyle \frac{1}{2}} \lVert \mathbb{A} \vec u - \vec{f} \rVert^2
			+ {\textstyle \frac{\lambda}{2}} \lVert \vec d^k - \mu \vec{u} - \vec{b}^{k} \rVert^2,\\
\label{eq2:dk+1}
\vec{d}^{k+1} &=& \arg \min_d \lVert \vec{d} \rVert_1
			+ {\textstyle \frac{\lambda}{2}} \lVert \vec d - \mu \vec{u}^{k+1} - \vec{b}^{k} \rVert^2,\\			
\label{eq2:bk+1}
\vec{b}^{k+1} &=& \vec{b}^k + \mu \vec{u}^{k+1} - \vec{d}^{k+1},
\end{eqnarray}
starting from $\vec{d}^0=\mathbf{0}$, $\vec{b}^0=\mathbf{0}$, and $\vec{u}^0=\mathbf{0}$. 
We use the conjugate gradient method to perform the $\ell_2$ minimization in Eq.~(\ref{eq2:uk+1}). 
The second step, Eq.~(\ref{eq2:dk+1}), separates into individual vector components and can be solved explicitly using shrinkage as
\begin{equation}
d^{k+1}_n = \text{shrink} \left( \mu u^{k+1}_n + b^{k}_n, 1/\lambda \right).
\end{equation}
The final step of the split Bregman cycle, Eq.~(\ref{eq2:bk+1}), adds
back the residual deficit in the $\ell_1$ term, in complete analogy
with the Bregman iteration Eq.~(\ref{eq:fk+1}). The results do not
depend on the value of the parameter $\lambda$, although an unsuitable
choice will lead to very slow or failed convergence. We find that in
practice an optimal $\lambda$ can easily be found from a few trial
runs at a fixed value of $\mu$, and then kept fixed for any $\mu$.
Just like FPC, the split Bregman iteration provides an exact solution
to the basis pursuit denoising problem Eq.~(\ref{eq:const_min_w_mu}),
but in contrast to the Bregman approach of Sec.~\ref{sec:Bregman},
small values of $\mu$ may be needed to solve the noiseless basis
pursuit problem Eq.~(\ref{eq:const_min}). In practice, we find that
the convergence rate of the split Bregman method is almost always
faster than those of the Bregman or FPC algorithms, and greatly so for
large, ill-conditioned sensing matrices.

\subsection{Choice of structures for CSCE}
\label{sec:structures}

An important practical question regards the best strategy for choosing
structures $\sigma$ to include in the training set. Mathematical
theorems from compressive sensing provide a definite answer to this
question. The key idea is the notion of coherence between the measurement and
representation basis. The representation basis $\Phi=\{\phi_j\}$ is
used to express the signal as a sparse series expansion (e.g., plane
waves form the representation basis for the Fourier series), while the
measurement basis $\Psi=\{\psi_k\}$ contains
all possible measurements. For the Fourier example in
Sec.~\ref{sec:Fourier}, the measurement basis is given by delta
functions, i.e., signal values at certain points in time. Assuming
that both $\psi_j$ and $\phi_k$ are normalized and orthogonal, the
coherence is defined as the maximum overlap between
them:\cite{donoho2001}
\begin{equation}
\label{eq:coh1}
\nu (\Phi,\Psi) = \sqrt{N} \max_{j,k} | \langle \phi_j, \psi_k \rangle |.
\end{equation}
In the Fourier example of Sec.~\ref{sec:Fourier}, the scalar products
are all $| \langle \phi_j, \psi_k \rangle | = N^{-\frac{1}{2}}$, which
corresponds to the lowest possible coherence, $\nu=1$. In contrast,
the highest possible value $\nu=\sqrt{N}$ would be obtained by
directly measuring the amplitudes of
the individual sinusoidal components of the signal, i.e., if plane
waves were chosen as the measurement basis functions. Coherence is key
in determining the number of measurements required to recover a given
sparse signal with $S$ nonzero components: the higher the coherence,
the higher the required number of measurements. More
quantitatively, the probability of correct signal recovery from $M$
measurements exceeds $1-\delta$ if the number of measurements
satisfies $M \ge C \nu^2 (\Phi,\Psi) S \log(N/\delta)$, where $C$ is a
constant and $S$ is the number of nonzero components;\cite{Candes,candes2006stable} a
similar result holds for compressive sensing in the presence
of noise. This expression shows that the worst possible strategy for
recovering sparse signals is to choose the same measurement basis as
the one used in sparse representation ($\nu (\Phi,\Psi) \approx
\sqrt{N}$), since this would require a number of measurements equal to
the number of unknown coefficients, $N$.

In cluster expansion, the representation basis are formed by
symmetry-distinct cluster types and the measurements are represented
by structures $\sigma$. The corresponding representation basis
functions are Kroenecker deltas, $\phi_{g} (f) = \delta_{fg}$, where
$f$ and $g$ are cluster numbers. The measurements are represented by
symmetry-inequivalent structures $\sigma$, and the corresponding basis
functions are given by normalized rows of the cluster correlation
matrix, i.e., $\psi_{\sigma} (f) = \bar\Pi_f (\sigma) / \sqrt{\sum_{f'}
  \bar\Pi_{f'} (\sigma)^2}$. The coherence is given by the maximum scalar
product between the two, which is
\begin{equation}
\label{eq:coh2}
\nu (\Phi,\Psi) = \sqrt{N} \max_{\sigma,f} \frac{| \bar\Pi_f (\sigma) |}{\sqrt{\sum_{f'} \bar\Pi_{f'} (\sigma)^2}}.
\end{equation}

Because random matrices with independent identically distributed (i.i.d.)
entries are incoherent with almost any representation basis,
they occupy a special place in compressive sensing. If the possible
measurements are designed by selecting $N$ uniformly distributed
random vectors on the unit sphere, followed by subsequent
orthogonalization, the coherence between $\Phi$ and $\Psi$ is on the
order of $\sqrt{2 \log N}$.\cite{candes2008introduction} 
This suggests the following simple strategy for
selecting structures for CSCE:
\begin{itemize}
\item Generate $M$ uniformly distributed random vectors $\psi_{\sigma}
  (f)$ on the unit sphere ($\sigma=1, \ldots, M$)
\item Orthogonalize $\psi_{\sigma} (f)$
\item Match each $\psi_{\sigma} (f)$ onto a real structure $\sigma$
  with normalized correlations $\bar\Pi_f (\sigma) / \sqrt{\sum_{f'}
    \bar\Pi_{f'} (\sigma)^2}$ approximating $\psi_{\sigma} (f)$ as closely
  as possible
\end{itemize}
The last step can be conveniently performed by enumerating all
possible ordered structures up to a certain size of the unit cell using the methods of 
Refs.~\onlinecite{hart2008algorithm,hart2009generating} and
then choosing the best matches from this list. We stress that the
somewhat counterintuitive strategy of selecting random structures
follows from the general mathematical properties of $\ell_1$-based
compressive sensing and represents the best possible method for
choosing structure sets for CSCE.

Parenthetically we note that selecting structures at random makes for a remarkably
  simple approach to generating input data. The ``structure selection'' problem, that is, deciding
  which structure to use to train the model, has been a vexing problem in the cluster expansion
  community since cluster expansions first began to be trained with first-principles data. At first,
  structures that were easy to calculate (few atoms per unit cell) were selected. In later years, more
  sophisticated approaches came to be used\cite{atat,seko2009cluster,ddj_fraction_factorial_design}, but a simple,
  easy-to-implement solution has remained elusive. Compressive sensing not only solves the ``cluster
  selection'' problem (because it makes unbiased selections from a huge set of clusters) but also
  overcomes the structure selection problem because it dictates that the best strategy is simply to
  select structures at random.

\subsection{Effect of noise and its relation to optimal $\mu$}
\label{sec:mu} 

The lone adjustable parameter, $\mu$, should be chosen
to achieve the optimal balance between the sparseness of the 
ECI's and the RMS fitting error for the training set. The
effect of $\mu$ on the calculated ECI's is most
transparently seen by analyzing the FPC equations
(\ref{eq:FPC1}) and (\ref{eq:FPC2}), which show that $\mu$ controls the
energy cutoff for the gradient of the $\ell_2$ norm of the residuals:
components of $\vec g$ with absolute values $|g_f| \le \mu$ will be
set to zero by the shrinkage operator and therefore will be
excluded from the model. In what follows, we show that
the optimal value for $\mu$ is proportional to the level of 
noise (random and systematic) in the training data. 

We first consider the relation between the normalized 
sensing matrix $\mathbb{A}$ in Eq.~(\ref{eq:FPC1}) 
and the CSCE correlation matrix $\bar\Pi$: they are related by
$\mathbb{A}=\bar\Pi/\sqrt{\alpha_{\bar\Pi}}$, where $\alpha_{\bar\Pi}$ is the largest eigenvalue
of $\bar\Pi\bar\Pi^T$. The corresponding relation for the 
measurement vectors is $\vec{f}=E/\sqrt{\alpha_{\bar\Pi}}$. 
The distributions of the extremal eigenvalues for
ideal random matrices are known from the theory of principal
component analysis.\cite{Johnstone2001} 
However, it is not immediately clear that the eigenvalue
distributions found for i.i.d. random matrices will be directly
applicable to the CSCE correlation matrices $\bar\Pi$
because the correlation values for real structures are neither 
independent nor identically distributed, and hence the entries 
of $\bar\Pi$ are only approximately i.i.d.  We have numerically 
calculated the distribution of the largest eigenvalue of 
$\bar\Pi\bar\Pi^T$ using subsets of $1 \le M \le 500$ fcc-based ordered
structures with 12 or fewer atoms in the unit cell.\cite{hart2008algorithm}
We considered $N=986$ correlations (up to six-body terms) and averaged 
the calculated eigenvalues over 1000 subsets randomly drawn from 
the above list of 10850 structures. We find that, for a fixed $N$, 
the average value of $\alpha_{\bar\Pi}$ increases linearly 
with the number of structures $M$. Therefore, 
$\mathbb{A} \propto \bar\Pi / \sqrt{M}$.

\begin{figure*}[ht]
\noindent
\parbox{1.0\linewidth}{
  \includegraphics[width = 0.295\linewidth]{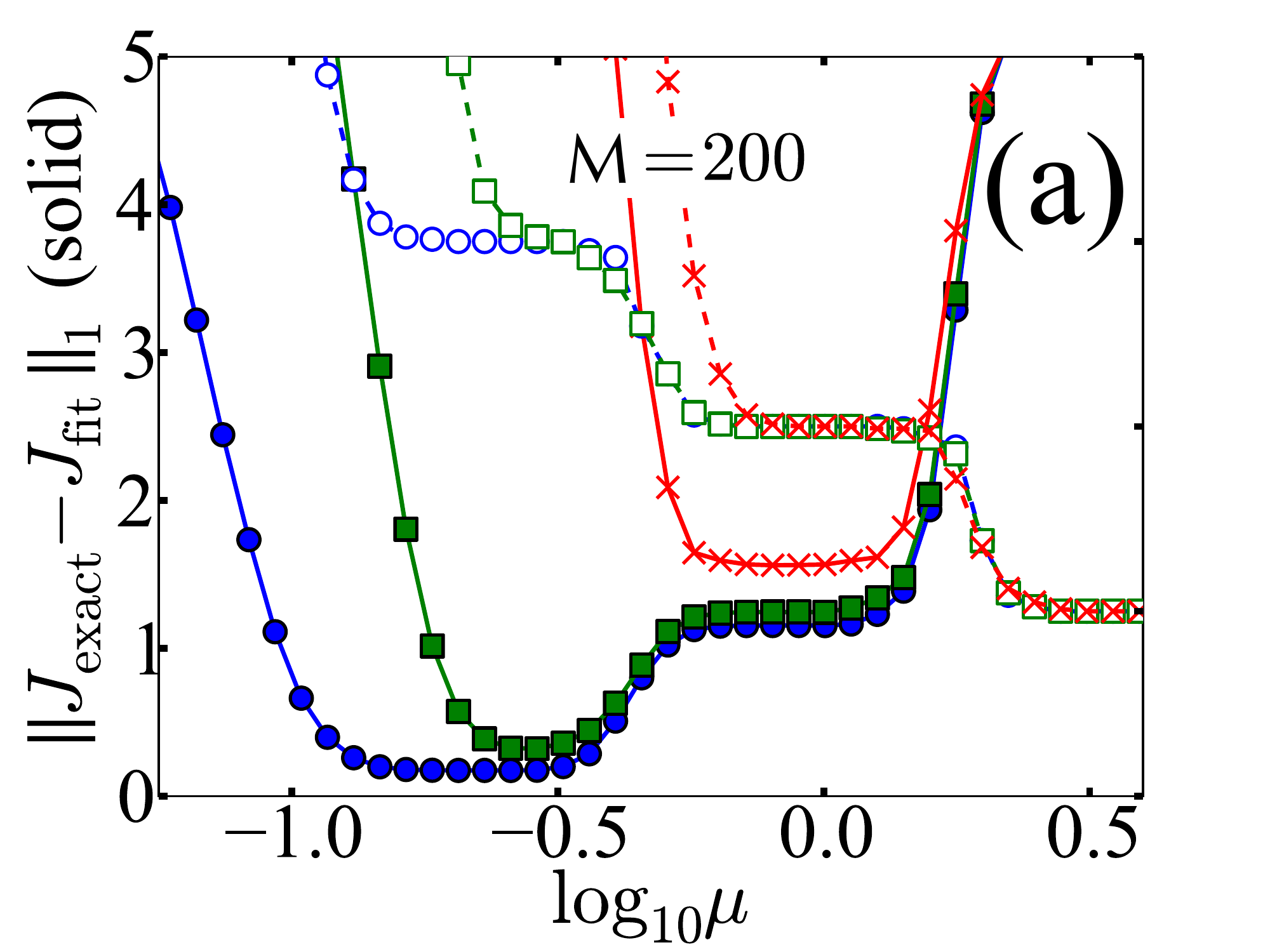}\hspace{-9mm}\includegraphics[width = 0.307\linewidth]{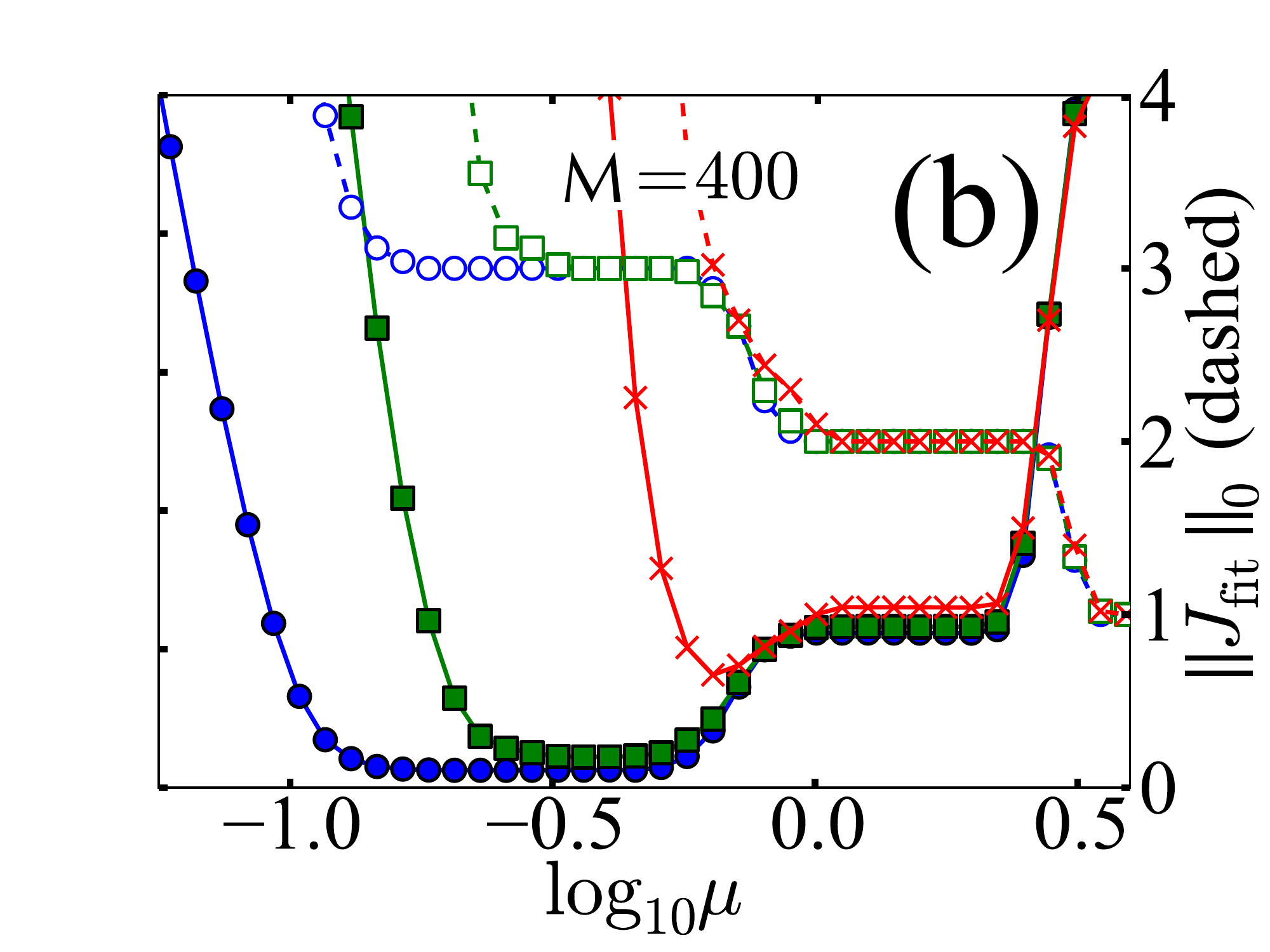}\hfill
\includegraphics[width = 0.4\linewidth]{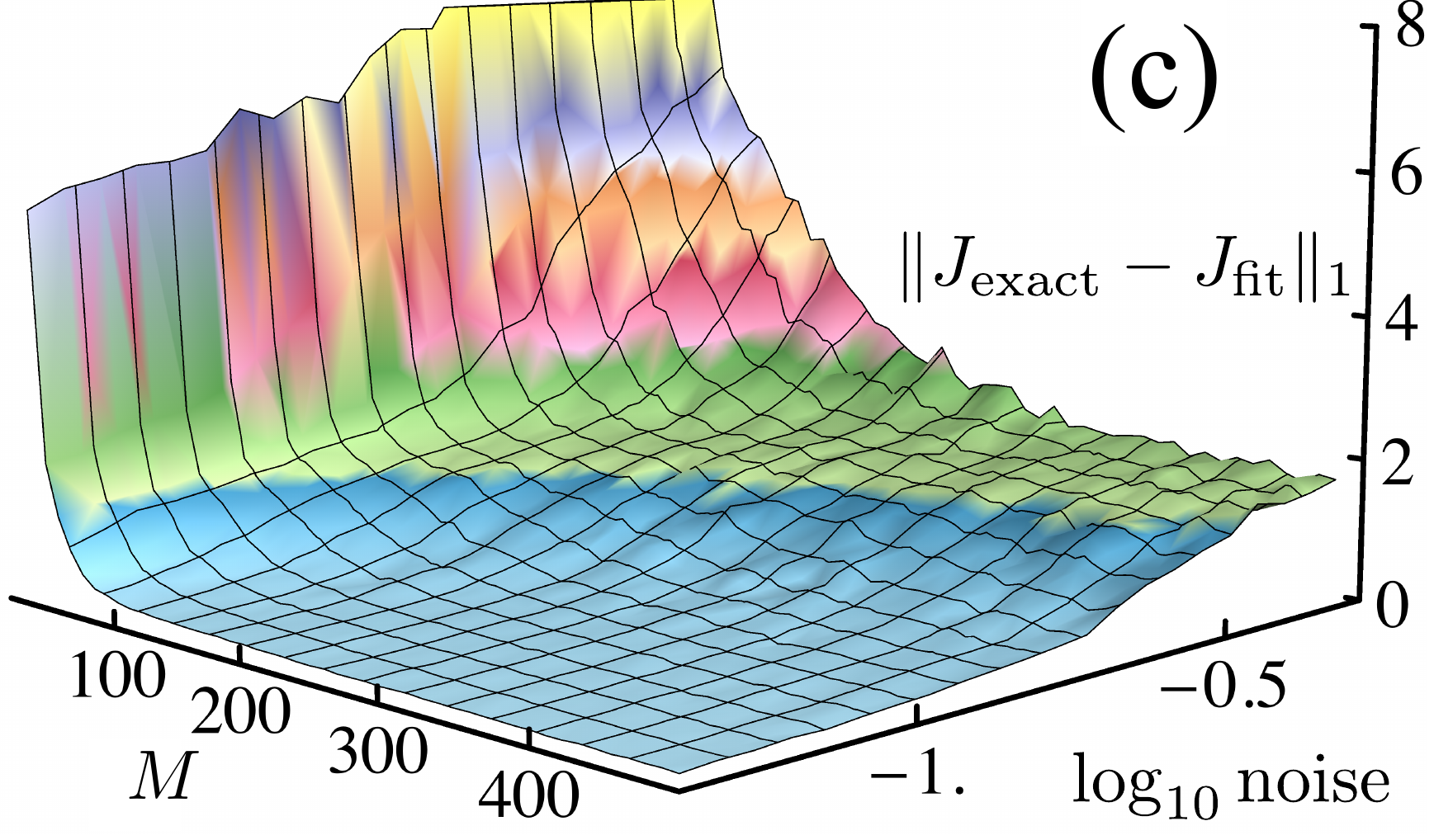}}
  \caption{\label{fig:toy_mu}  $\lVert J_\text{exact}
    -J_\text{fit} \rVert_1$(solid) and $\lVert J_\text{fit}
    \rVert_0$(dashed) vs log$_{10}\mu$ for the short-ranged pair model with $M =
    200$ (a) and $M = 400$ (b). Random uniform noise of
    $\sim 10\%$(blue circles), $20\%$ ( green squares), and $50\%$ (red
    ``x''s) of the noiseless energies was added to the fitting
    structures. (c) $\lVert J_\mathrm{exact} -J_\mathrm{fit} \rVert_1$
  vs the number of fitting structures and the noise level.  Each point
represents an average over $\sim$100 different subsets of $M$ structures.
}
\end{figure*}

{\it Random noise:\/} Here we demonstrate
that CSCE is not only stable with respect to noise in the
input data, but that it can also filter out the effects 
of noise on the calculated ECI's. We assume that the DFT 
formation energies $E(\sigma)$ contain random noise which is
represented by a vector $\vec{\eta}_\epsilon$ 
of length $M$ and i.i.d random components with variance 
$\epsilon_\text{rand}^2$. The contribution of  $\vec{\eta}_\epsilon$ 
to the FPC gradient in Eq.~(\ref{eq:FPC1}) is  given by
\begin{equation}
\label{eq:noisegrad}
\delta g_f \propto - \frac{1}{M} \sum_{\sigma=1}^M \bar\Pi_f (\sigma) \eta_\epsilon (\sigma),
\end{equation} 
where the factor $1/M$ comes from the fact that both the sensing matrix
$\mathbb{A}$ and the measurement vector $\vec{f}$ are related to the correlation
matrix $\bar \Pi$ and input energies $E$ by a normalization factor $1/\sqrt{\alpha_{\bar\Pi}}$. 
If the structures are chosen randomly according to the prescription outlined in Sec.~\ref{sec:structures}, 
then $\bar\Pi_f (\sigma) \in [-1,1]$ are approximately i.i.d. Hence, the individual terms under the 
summation sign in Eq.~(\ref{eq:noisegrad}) will be randomly distributed 
with a mean of zero and a variance proportional to $\epsilon^2_\text{rand}$. 
To deduce the behavior of $\delta g_f$ in the limit of large $M$, 
one can apply the central limit theorem (CLT) of classical statistics, 
which states that the average of $M$ random terms is normally 
distributed with a variance that is given by the variance of the individual 
terms divided by $M$, i.e., the variance of $\delta g_f$ is proportional
to $\epsilon^2_\text{rand}/M$.
It then follows from the properties of the normal distribution that the 
average $\ell_1$ norm of the noise term in the gradient decreases
with the size of the training set as
\begin{equation}
\label{eq:noisegrad2}
||\delta g_f||_1 \propto \frac{\epsilon_\text{rand}}{\sqrt{M}}.
\end{equation}
This relation demonstrates an important noise-tolerance aspect of 
CSCE, which guarantees that the true physical ECI's will be recovered 
even if the training data sets contains uncorrelated random noise of arbitrary
magnitude, provided that the number of data points is sufficiently large. 
The practical significance of this feature cannot be overstated: not only
is CSCE stable with respect to random noise, but an absolute numerical
accuracy in the DFT energies is not even needed to recover the correct 
ECI's!\footnote{This applies only to random numerical errors in the DFT 
formation energies and excludes systematic errors, such as those due 
to the approximate nature of the exchange-correlation functionals.} 
Equation~(\ref{eq:noisegrad2}) also offers guidance for
choosing $\mu$ to smooth the effect of random noise:
as long as $\mu \simeq ||\delta g_f||_1$, the 
contribution of noise to the gradient will be zeroed out in the 
shrinkage step [Eq.~(\ref{eq:FPC2})] and will not affect the 
calculated ECI's. In practice, however, the optimal value of $\mu$ is difficult to 
determine using Eq.~(\ref{eq:noisegrad2}) because the level of 
noise in the DFT formation energies is not known {\it a priori,\/}
and approaches based on optimizing the predictive error
are more practical.

{\it Systematic noise:\/} We next consider the effect of 
systematic noise due to errors in the ECI's, which we denote by $\delta J_f$. 
These errors contribute a term 
$\delta E(\sigma) = \sum_f \bar\Pi_f (\sigma) \delta J_f $ to the residual,
and the corresponding error in the FPC gradient is given by
\begin{equation}
\label{eq:sysnoisegrad}
\delta g_f \propto - \sum_{f'} \langle \bar\Pi_f \bar\Pi_{f'} \rangle \delta J_{f'},
\end{equation} 
where we have introduced a correlation matrix for cluster correlations 
$\bar\Pi$ calculated over the training set:
\begin{equation}
\label{eq:Picorr}
\langle \bar\Pi_f \bar\Pi_{f'} \rangle = \frac{1}{M} \sum_{\sigma=1}^M \bar\Pi_f (\sigma) \bar\Pi_{f'} (\sigma).
\end{equation}
This matrix is of fundamental importance for CSCE because it describes how
the value of one ECI is affected by errors in the other ECI's, or the degree of cross-contamination
between systematic ECI errors. Minimum sensitivity to cross-contamination is achieved 
when $\langle \bar\Pi_f \bar\Pi_{f'} \rangle$ is diagonal, but the latter case is impossible
to realize in practice due to the fact that there are rather pronounced correlations between the cluster averages 
in real structures. In the best case scenario, the correlation matrix  $\langle \bar\Pi_f \bar\Pi_{f'} \rangle$
will be approximately diagonal if the training set structures are chosen 
randomly according to the algorithm proposed in Sec.~\ref{sec:structures}. 
Indeed, if the average cluster correlations $\bar\Pi_f (\sigma)$ 
are approximately i.i.d., the off-diagonal elements of the correlation matrix 
$\langle \bar\Pi_f \bar\Pi_{f'} \rangle$ tend to zero with increasing $M$, 
while the diagonal elements remain $O(1)$:
\begin{equation}
\label{eq:Picorr2}
\langle \bar\Pi_f \bar\Pi_{f'} \rangle \, = \, 
\begin{cases}
	\langle \bar\Pi_f^2 \rangle \; \text{for } f=f' \\
	O \left( \frac{1}{\sqrt{M}} \right) \; \text{for } f \neq f' 
\end{cases}.
\end{equation}
Hence, in the limit of large $M$, CSCE based on a randomly chosen training set 
cleanly separates the contributions of the systematic ECI errors to the gradient, i.e., 
the ECI error for cluster $f$ only affects the 
component $f$ of the gradient, enabling accurate recovery of the 
correct solution. This is an important feature for any physics model-building 
approach because it guarantees the stability of the solution with respect
to the interactions that are not represented within the chosen basis set.
Furthermore, these considerations offer another insight into the
physical meaning of the parameter $\mu$: it can be used to filter out the 
cross-contamination due to effects of systematic noise if chosen as
\begin{equation}
\label{eq:musyst}
\mu \sim \frac{||\delta \vec J||_\infty}{\sqrt{M}},
\end{equation}
where $||\delta \vec J||_\infty$ is the magnitude of the largest error
in the cluster interactions. Since the diagonal contribution to the gradient remains constant with 
increasing $M$, successively smaller ECI's can be extracted by increasing 
the size of the training set $M$ and simultaneously decreasing the value of $\mu$
according to Eq.~(\ref{eq:musyst}). Unfortunately, the practical 
value of this expression is limited because the ECI errors are not known,
and approaches based on minimizing the prediction errors or CV scores 
should be used instead.

The preceding analysis shows that  $\mu$ can be interpreted as a parameter
controlling the filtering of the noise in the calculated energies, including both
random noise due to numerical errors in the DFT formation enthalpies
and systematic noise due to cluster interactions that are not
recoverable using the given structure set.
Expressing the total noise level as a sum of random and systematic contributions, 
$\epsilon^2 = \epsilon^2_\text{rand} + \epsilon^2_\text{syst}$, 
the effect of both is expected to decrease the inverse of the size of the training set,
and the optimal value of $\mu$ is expected to vary as $1\sqrt{M}$.
We note here that the Bregman and split Bregman iterations contain additional 
noise-filtering steps [Eqs.~(\ref{eq:fk+1}) and (\ref{eq2:bk+1})] which 
add back the residual to the residual of the next iteration.
As a result, the optimal value of $\mu$ will in general vary between 
the different $\ell_1$ optimization approaches, even though the solutions
and the predictive errors are practically the same.\cite{yin2008bregman,goldstein2009split}

\section{Applications}
\label{sec:applications}

\subsection{Short-ranged pair model with noise}
\label{sec:toymodel}

We first work with an ad-hoc cluster expansion example where we choose
a set of sparse coefficients and then use them to compute the energies
of various crystal structures for use as input to CSCE. The advantage of
this approach is that knowing the exact solution {\it a priori\/}
allows us to easily determine the accuracy of the solution found by CSCE
and determine how numerical noise influences the performance of the
algorithm.  While this example is certainly not representative of any
real alloy system, it clearly illustrates some key features of the
method, particularly how CS performs with noisy data.

Using the \textsc{uncle} \cite{lerch2009uncle} framework the following
clusters on an fcc lattice were enumerated: 141 pairs, 293 triplets,
241 four-bodies, 87 five-bodies, and 222 six-bodies (986 clusters in
total, including the onsite and empty clusters). The coefficients of the three
shortest nearest-neighbor pairs were chosen as 10, 4, and 1,
respectively; all other coefficients were set to zero. Uniformly
distributed random noise equal to $\sim$ $10\%$, $20\%$, and $50\%$ of
the noiseless energies was added to the computed energies $E(\sigma)$. 
We emphasize that these noise levels \emph{significantly} exceed typical 
numerical errors in the calculated formation enthalpies 
from state-of-the-art quantum mechanics codes.\footnote{In our estimation,
numerical errors in the calculated DFT formation energies 
are only a few meV/atom for the case of  Ag-Pt compounds 
considered in Sec.~\ref{sec:AgPt}.}

The values of each of
the 986 basis functions were computed for all structures in the
training set, thus forming the sensing matrix, $\mathbb{A}$.
The rows of the sensing matrix, $\mathbb{A}$, which each
represent a training set structure, were constructed by drawing randomly
from a uniform distribution on $[-1,1]$. For real systems,
such as Ag-Pt in the next section, these rows should be mapped 
onto real crystallographic configurations as described in 
Sec.~\ref{sec:structures}. 
However since the quality of the fit for the short-ranged pair case 
was found to be unaffected by this mapping, either favorably or adversely, 
we chose to simply use the random vectors themselves in order to simplify
computations. 

Figures \ref{fig:toy_mu}(a) and \ref{fig:toy_mu}(b) illustrate the
performance of CS by showing two quantities: 1) the $\ell_1$-norm of
the difference between the exact and fitted coefficients 
($\lVert
J_\text{exact}-J_\text{fit} \rVert_1$
), and 2)
the number of non-zero coefficients ($\ell_0$-norm of the solution,
$\lVert J_\text{fit} \rVert_0$). We varied $\mu$ to investigate
it's optimal values for a given noise level.  Each data point in
Fig.~\ref{fig:toy_mu} was obtained by averaging over approximately 100
different sets, each of size $M=200$ or $400$.
 
The curves in Fig.~\ref{fig:toy_mu} exhibit a series of plateaus, each
one indicating a region over which the extracted solution remains
practically unchanged. Notice, for example, the plateau located
between $\log_{10}\mu = -0.75$ and $\log_{10}\mu = -0.4$ in the
$\lVert J_\text{fit} \rVert_0$ {\it vs.\/} $\mu$ curve for $M = 200$ and the
lowest noise content (circle markers). This plateau indicates that CSCE
has extracted three non-zero coefficients. Furthermore, the value of
$\lVert J_\text{exact} -J_\text{fit}\rVert_1$ drops close to zero in this
range, indicating that CSCE has found essentially the exact answer.
Using values of $\mu$ below the optimal range results in sharp
increases in both the number of nonzero coefficients and in the error
$\lVert J_\text{exact} -J_\text{fit}\rVert_1$, indicating overfitting.

Conversely, $\mu$ values above the optimal range result in fewer
non-zero coefficients and an incremental increase in $\lVert
J_\text{exact} -J_\text{fit}\rVert_1$, probably indicating underfitting. As a
function of increasing $\mu$, one first obtains a plateau where the CS
reproduces the two largest expansion coefficients (10 and 4), followed
by another plateau where only the largest coefficient is reproduced.
This example illustrates the important point that CS is largely \emph{insensitive} to
the choice of $\mu$---the ability to recover the correct solution does not depend
on the exact value of $\mu$, as long as it lies within an
optimal, but broad, range. 

Upon increasing the noise in the fitting data at a fixed data set size
[compare the curves marked by circles and squares in
Fig.~\ref{fig:toy_mu}(a)], the plateaus in $\lVert J_\text{fit}
\rVert_0$ vs $\mu$ become narrower until the highest plateau,
corresponding to full recovery of the true solution, disappears
completely (``x'' markers in Fig.~\ref{fig:toy_mu}).  At the same
time, the minimum in the error $\lVert J_\text{exact}
-J_\text{fit}\rVert_1$ vs. $\mu$ is increasing incrementally.  This
displays the robustness and stability of CS---even at a very high
noise level we are able to recover the majority of the signal content.

The shift towards higher values 
of optimal $\mu$ upon increasing noise level in 
Fig.~\ref{fig:toy_mu} is consistent with the physical 
interpretation of $\mu$ as the threshold for noise filtering 
given in Sec.~\ref{sec:mu}. We also note that an
increase in the number of structures $M$ tends to 
slightly lower the optimal $\mu$, which can be attributed 
to a fuller recovery of the correct solution and an associated
decrease in the systematic noise.

Figure \ref{fig:toy_mu}(c) displays $\lVert J_\text{exact} -J_\text{fit}
\rVert_1$, averaged over approximately 100 random subsets, as a function of
$M$, the number of fitting structures, and the noise level.
Here we see the same plateau structure
found in Fig. \ref{fig:toy_mu}(a), with the lower (blue) plateau indicating
essentially an exact fit.  This plot demonstrates that, for all noise
levels considered (up to as high as 50\% of the noiseless energies!),
there remains a training set size for which the
exact solution will be recovered. 
    
\subsection{Actual alloy example: Ag-Pt}
\label{sec:AgPt}

\begin{figure}
\centering
\includegraphics[width = 1\linewidth]{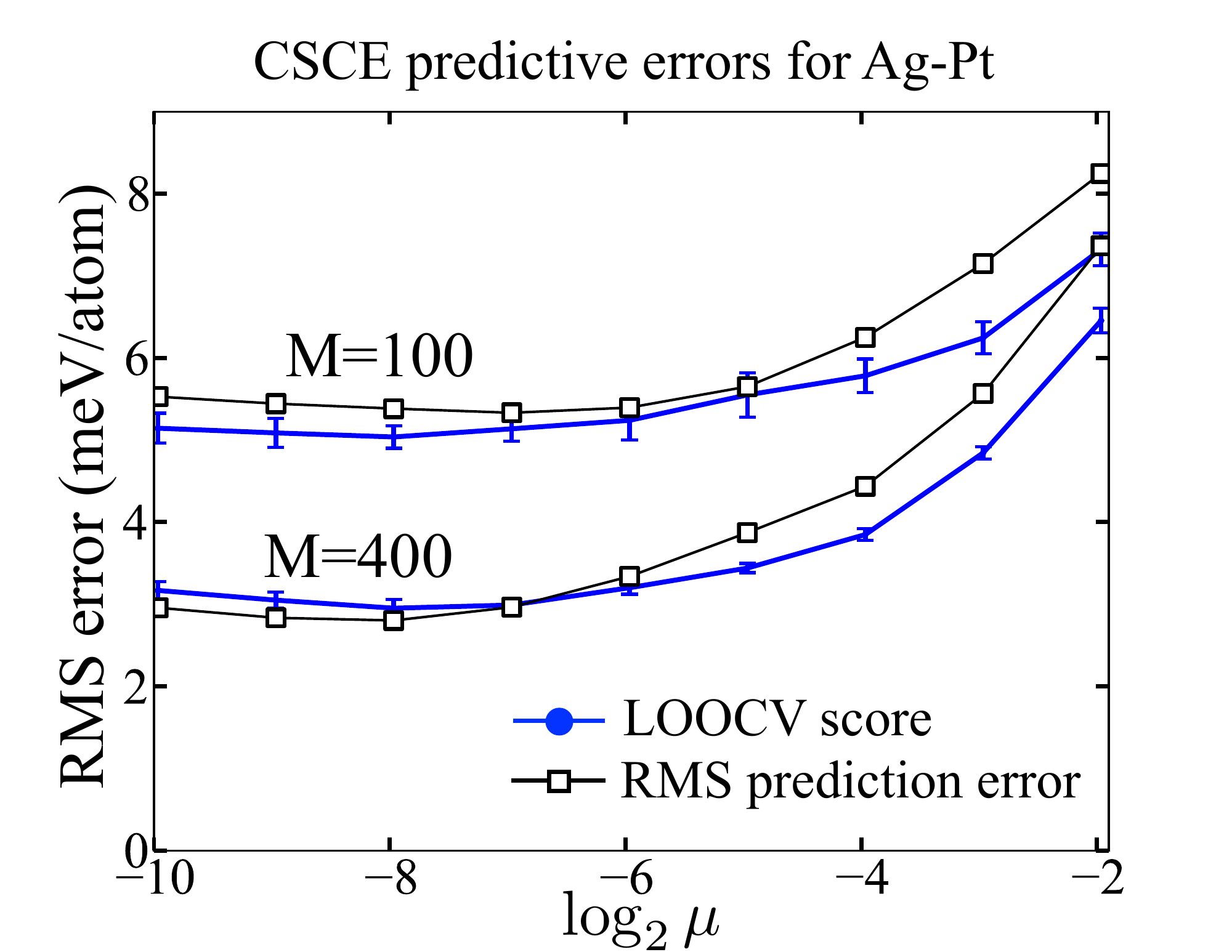}
\caption{Root-mean-square errors for the prediction set (black
  line with empty squares) and the leave-one out cross-validation
  score (LOOCV, solid blue line) as functions of the parameter
  $\mu$. LOOCV has been averaged over 10 randomly drawn sets of 100 (400)
  structures, and the error bars were calculated from the variance in
  the predicted LOOCV scores over these sets. Predictive errors for
  the hold-out set and the fitting errors for the training set were
  averaged over 500 different sets of 100 (400) structures; the corresponding
  error bars are smaller than the size of the symbols.}
\label{fig:mu}
\end{figure}

Having explained the basic properties of CSCE for a model system, we now
test its performance on real DFT data for binary Ag-Pt alloys on a
face-centered cubic (fcc) lattice. Ag-Pt was chosen due to a report of
unusual ordering tendencies\cite{Durussel1996} which are non-trivial
to reproduce with current state-of-the-art CE methods. The energies of 
more than 1100 Ag-Pt fcc-based crystal structures\footnote{Such a 
large number of structures was only chosen to test the performance 
of different CE methods and is several times larger than typical training 
set sizes used in state-of-the-art CE methods.}
were calculated from the density-functional theory (DFT) using the
\textsc{vasp} software.\cite{Kresse:1999wc,kresse1996efficiency} 
We used projector-augmented-wave (PAW)
potentials\cite{Blochl:1994dx} and the generalized gradient approximation 
(GGA) to the exchange-correlation functional proposed by 
Perdew, Burke and Ernzerhof.\cite{PBExc}
To reduce random numerical errors,  equivalent $k$-point meshes were 
used for Brillioun zone integration.\cite{froyen1989brillouin} Optimal choices of
the unit cells, using a Minkowski reduction algorithm, were adopted to
accelerate the convergence of the calculations.\cite{Minkowski}
The effect of spin-orbit coupling was not included in our calculations because it's effect
was shown to be a simple tilt of the calculated energies, as explained 
in Ref.~\onlinecite{nelson2012ground}.

Out of a total of approximately 1100 structure energies calculated for 
this system, 250 were chosen at random to be held out of the fitting process and
used for prediction. This ``holdout'' set remained unchanged for all
fitting sets chosen.  Of the remaining 850 data points available for
fitting, subsets of up to $N=400$ were chosen to be used as 
CSCE training data.

We start by illustrating the performance of two different methods 
for selecting the
optimal value of $\mu$. First, we varied $\mu$ and calculated the
standard LOOCV score over 10 different randomly drawn subsets 
of $M$ structures; the results are shown by blue curves in 
Fig.~\ref{fig:mu}. It is seen that the LOOCV scores reach their 
minima at $\mu \approx 4$ and 2 ~meV/atom for $M=100$ and 400, 
respectively, which we interpret as the optimum $\mu$'s providing maximum 
predictive power. Second, we calculated the average prediction 
errors for all structures left out of the fitting set, which are 
represented by the black dotted lines in Fig.~\ref{fig:mu}. 
We see that the RMS errors for the prediction set 
largely follow the same behavior as the LOOCV scores, 
reaching minima at nearly identical $\mu$ values.

As expected, fitting errors for the training set (not shown here) decrease monotonically 
with decreasing $\mu$ and are significantly smaller than either the 
LOOCV scores or prediction errors for the hold-out set. The leveling off in both the
prediction errors and the LOOCV score at small values of $\mu$ can be
explained by noting that CSCE fits the training set perfectly and
further decrease of $\mu$ does not bring about noticeable changes in the
calculated ECI's. We note that this behavior is different from the short-ranged pair
model in the previous section, where decreasing $\mu$ below the
optimal range caused a rapid deterioration in the accuracy of the
calculated ECI's. We attribute this difference to the lower level of
noise in the Ag-Pt case, so that the range of $\mu$'s that leads to
acceptable ECI's is much wider than at the 20-50\% noise level
for the short-ranged pair model.

To compare the performance of CSCE with other
established methods, a discrete optimization (DO) scheme
as implemented in the state-of-the-art ATAT software 
package,\cite{vandeWalle2002,vandeWalle2009} was used. 
Note that the ATAT program is capable of employing advanced
algorithms beyond minimization of the LOOCV score to ensure 
that the ground state line is reproduced correctly. 
In order to make a straightforward comparison between CSCE and DO, 
we only used the LOOCV-based DO functionality of ATAT. 
Since the DO method for $N=986$ clusters on a training set 
of a few hundred structures takes several days to complete, averages were taken
over only 10 training sets of size $M$ 
(except for $M=400$ when we used 42 different training sets to perform 
statistical analysis of the calculated ECI's).  In order to simulate
building a complicated unknown model, we deliberately avoided applying
physical intuition (e.g., picking short-range interactions) and simply
performed the optimizations with minimal restrictions. 
The maximum number of reported ECI's was capped to $M/4$ for 
ATAT-based DO.  For CSCE, we used a fixed $\mu=8$~meV/atom 
and computed solutions for $500$ randomly chosen training sets of 
$M$ structures.

\begin{figure}
\includegraphics[scale = 0.415]{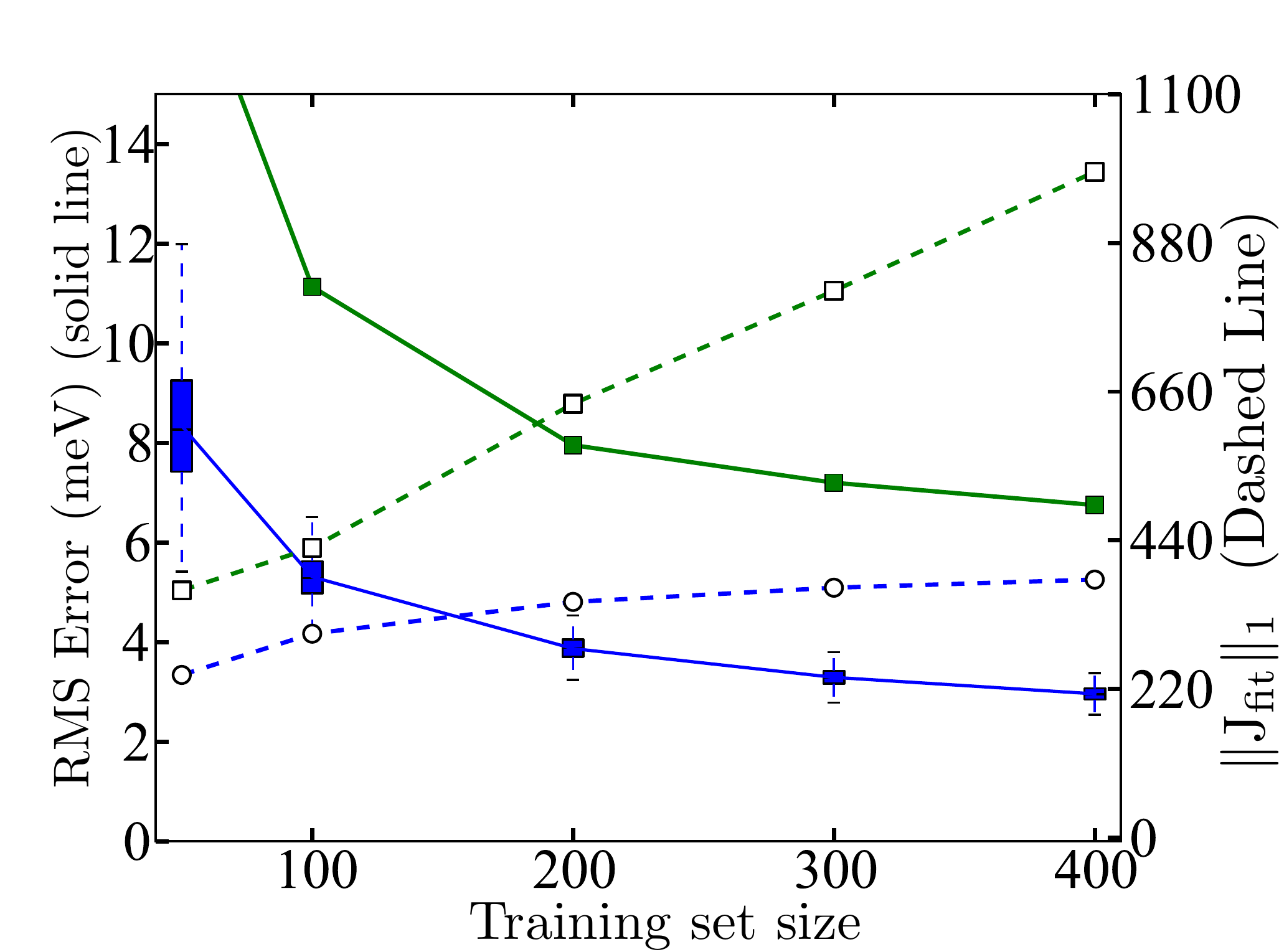}
\caption{Results from compressive sensing and leave-one-out
  cross-validation for the fcc-based, Ag-Pt alloy system.  The solid
  line gives the root-mean-square (RMS) errors for predictions made on
  a constant holdout set for CS(box and whisker) and leave-one-out cross-validation
  (squares). The dashed lines give the $\ell_1$-norm of the solution
  vector for both methods. }
\label{fig:cs_vs_cv}
\end{figure}

Figure \ref{fig:cs_vs_cv} shows a box and whisker plot of the RMS 
errors over the prediction set for CS solutions and
the mean RMS values for the DO solutions (box-and-whiskers were not
used for DO solutions due to the small number of DO fits). Each
box and whisker represents RMS values for approximately $500$
different fits.  We see that CSCE achieves an RMS error value much lower
(2.8~meV/atom) than LOOCV-based DO (6.8~meV/atom).
Furthermore, Fig.~\ref{fig:cs_vs_cv} shows that the $\ell_1$ norm of 
the solution increases almost linearly for the DO fit, while it levels off for
the CSCE fit, indicating that the latter is converging towards a stable solution,
while the former keeps adding large ECI's, a behavior suggestive of over-fitting.

\subsection{Statistical analysis of Ag-Pt ECI's}
\label{sec:statanalysis}

\begin{figure*}
\noindent
\parbox{1.0\linewidth}{
\includegraphics[width = 1.0\linewidth]{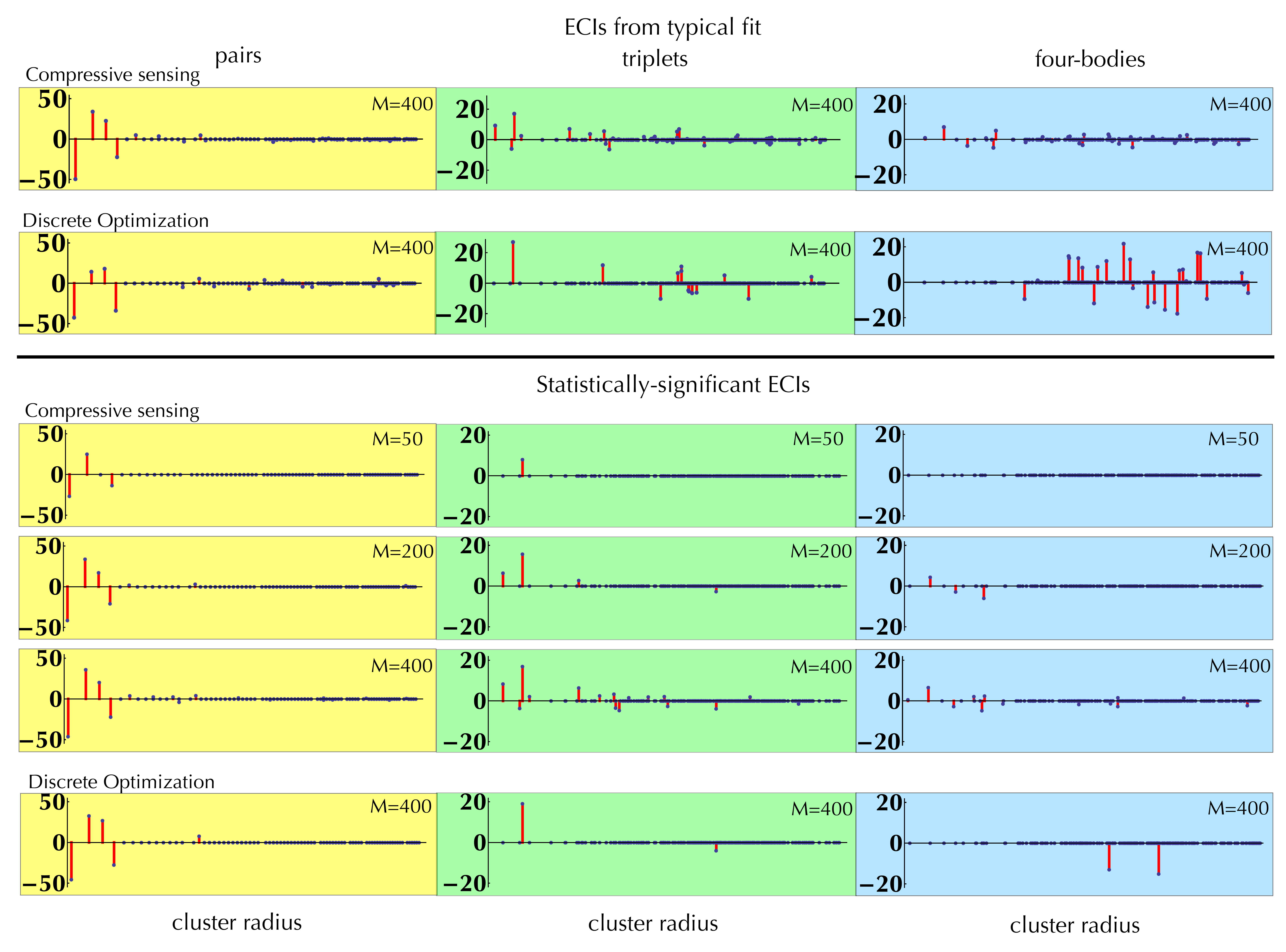}}
\caption{Comparison of the interaction coefficients found using the
  DO method implemented in ATAT software and compressive sensing.  
  The upper pane shows a
  comparison of two typical fits from CS and ATAT.  The lower pane
  shows the coefficients that were found to be statistically relevant
  from both methods.  The x-axis is the cluster radius, which is
  defined as the average distance from the center of mass of all
  cluster vertices. (Blue dots were placed on the x-axis even for
  clusters not found to be relevant to help the reader know the
  ordinal number of the relevant clusters.) Physical intuition suggests that
  shorter-radius, fewer-vertex clusters are the most important
  contributors in alloy energetics. Pair interaction coefficients
  found by both methods are similar.  As the number of
  vertices increases, CS finds coefficients in harmony with physical
  intuition, while DO finds spurious, long-ranged three- and
  four-body interactions. CS solutions also demonstrate a
  convergence to one specific solution as the size of the fitting set
  increases. (note: Triplets and quadruplets are shown on a scale from
-20 to 20 meV, different from the scale used for the pairs.)}
\label{fig:Js}
\end{figure*}

Because CSCE is fast, thousands of fits for many different training sets
can be computed in a few minutes. The results of all these fits
can be analyzed statistically to determine which coefficients are
consistently identified as contributors and to eliminate artifacts
due to a particular choice of the training set. This functionality, the
ability to gather enough data in a reasonable amount of time to
perform statistical analyses, is a significant advantage of CSCE 
over (slower) DO methods that can be used to gain insight into 
the probability distributions for the cluster interactions. 
These distributions can be used to quantify the uncertainty in the
CSCE predictions for physical properties that go beyond a 
simple LOOCV score or an RMS prediction error.
For instance, one can draw ECI's from the calculated
distributions and generate ground state convex hulls 
with statistical error bars on each structure,
quantifying the uncertainty in the predicted
$T=0$~K phase diagrams.

CSCE fits for $500$ different
fitting set choices were computed for Ag-Pt. Most of the resulting distributions
had only one sharp peak at zero, indicating that, independently of the choice of the training set, 
they were almost never selected by CSCE and therefore should be set to zero. 
Several ECI's exhibited a unimodal distribution with nonzero mean,
which were interpreted as strongly significant nonzero interactions. 
Finally, a fraction of the ECI's showed bi-modal distributions with two peaks of 
comparable weight and one of the peaks centered at zero energy. 
Since the latter ECI's were selected by CSCE with an approximately 50\% 
probability, they belong to the class of ``marginal'' interactions which were 
counted as significant only if their distribution mean was greater than 
one standard deviation.  To make a fair comparison between CSCE and the 
DO method implemented in the ATAT program, the same statistical 
criteria for determining relevant coefficients was used for the DO fits,
 even though data for only $42$ fits were available.

Figure \ref{fig:Js} gives a comparison of the CS-determined
coefficients and those found by DO. The upper pane compares a
typical DO fit with a typical CSCE fit, while the lower pane gives a
comparison of statistically relevant ECI's from both methods. 
The CSCE-derived ECI's appear to evolve towards one specific solution as
the size of the fitting set increases, indicating convergence of the
solution. Notice also
that the magnitudes of the CSCE coefficients decrease as the spatial
extent of the cluster increases and as the number of cluster
vertices increases (note that triplets and quadruplets are shown on a
scale from -20 to 20 meV, as opposed to -50 to 50 meV for pairs). 
This is in harmony with long-standing claims in
the CE community, and it confirms that a stable solution has been
found. DO-determined clusters follow this pattern for pair clusters
only. At higher vertex numbers, a typical DO fit finds non-physical,
spurious coefficients for three- and four- body interactions.
The set of statistically-relevant DO coefficients appear to be
lacking several important interactions, specifically short-ranged
three- and four-body interactions. This indicates that: (i) current
DO methods are much too slow to be able to gather enough 
statistics to do a meaningful statistical analysis, 
and/or (ii) current DO methods are very sensitive to the choice 
of the training set and fall short in their ability to identify physically 
relevant interactions without user guidance.

\begin{figure}
\includegraphics[width = 1.0\columnwidth]{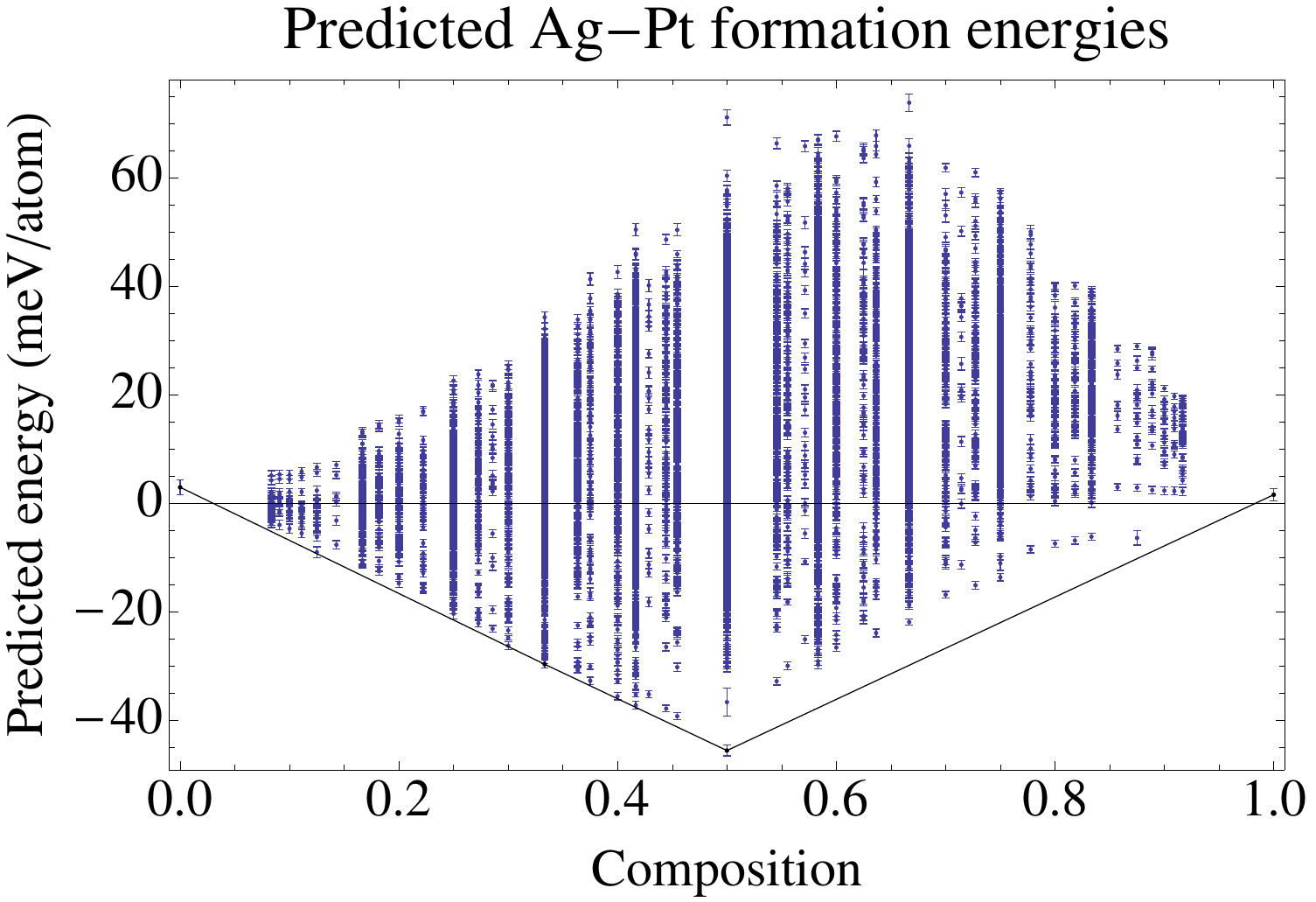}
\caption{Predicted CSCE formation energies obtained using the ECI's
  shown in Fig.~\ref{fig:Js}; error bars are standard deviation due to
  different random choices of $<=400$ structure subsets. Black solid
  line denotes the convex hull calculated from the average energies;
  only Ag, Ca$_7$Ge-type Ag$_7$Pt (barely, with a depth of less than 1
  meV/atom), $L1_1$ AgPt, and Pt are predicted to be $T=0$ K ground
  states.}
\label{fig:convexhull}
\end{figure}

Figure~\ref{fig:convexhull} shows the results of a ground state search
performed by using the statistically significant $M=400$ coefficients 
to predict the energies of all fcc-based superstructures up to 
12 atoms. Error bars were calculated from randomly 
drawn sets of $M=400$ structures. The ground state line in this 
figure is consistent with first-principles data for this system, 
which finds the same ground states as in Fig.~\ref{fig:convexhull}, 
with a few degenerate structures lying on the convex hull 
between $c=0.4$ and $0.5$.

This example shows that, in comparison with traditional cluster
selection methods, CS is not only simpler and faster (less than a
minute on a single CPU for CS versus \emph{days} for LOOCV at $M=400$),
but also produces more physical solutions that result in a 
significant improvement in physical accuracy.

\subsection{Protein folding application}
\label{sec:proteinCE}

\begin{figure}[t]
\includegraphics[scale =0.44]{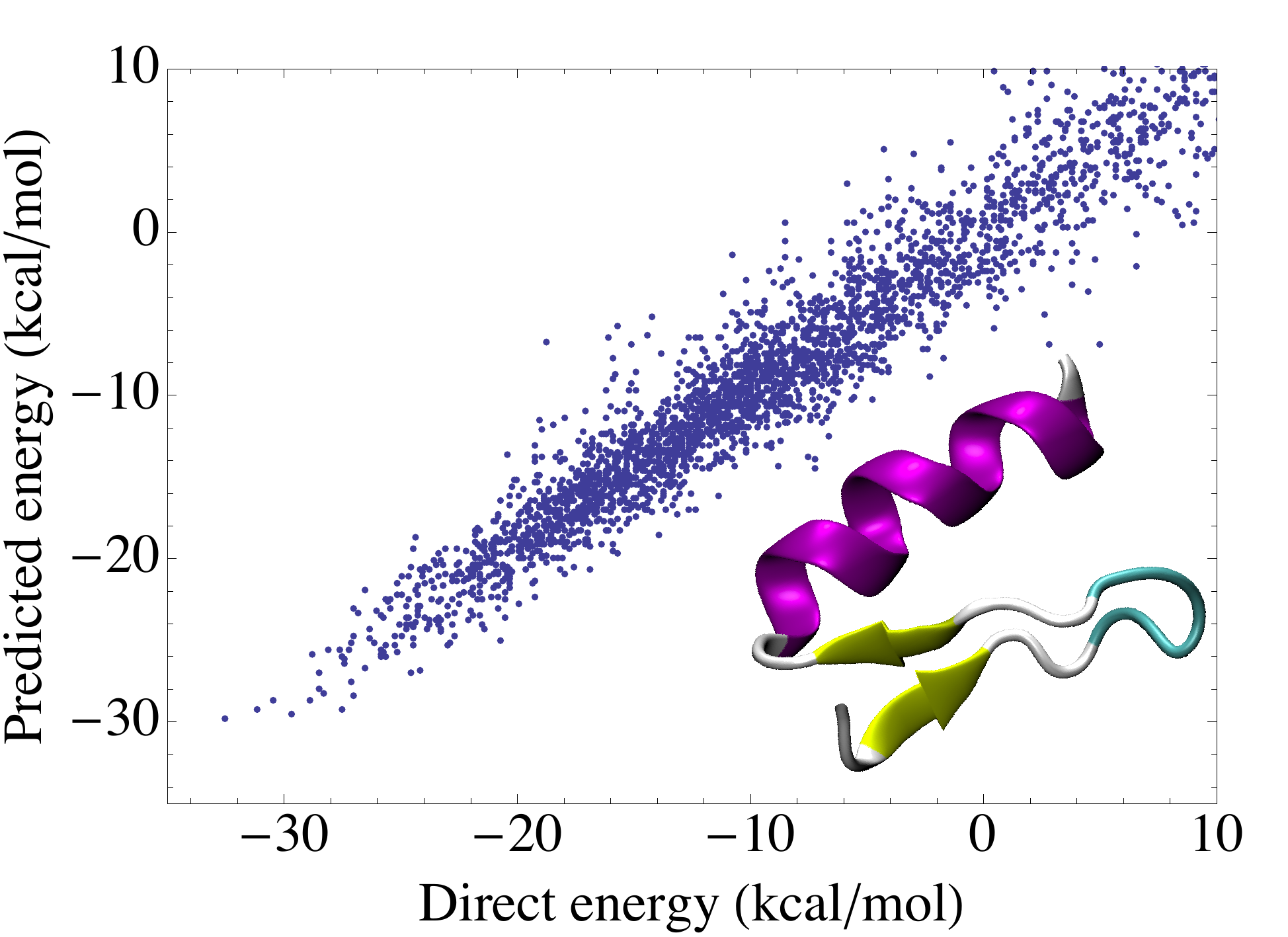}
\caption{ Predictive performance of CS for protein energetics in 
the zinc-finger structure (shown in the inset).}
\label{fig:zinc-finger}
\end{figure}

We now turn to a technically much more challenging case---that of
protein design in biology. Modeling the protein folding energies in
the zinc-finger motif represents a technically difficult test case
with novel applications in
biology.\cite{zhou2005coarse,grigoryan2009design} One of the key
problems in protein design is to find the sequence of amino acids
(AAs) which stabilizes a particular 3D structure, or {\it
  folding}. Physics-based energy functionals are considered to be some
of the most-promising methods in protein design since they link the
stability of the the folded 3D structure to the total free energy,
accurately accounting for electrostatics, van der Waals interactions,
and solvation effects. However, their use is problematic due to the
astronomical number of possible AA sequences for even very short
proteins. It was shown \cite{zhou2005coarse,grigoryan2009design} that
the CE model can be generalized to describe protein energetics,
allowing very fast direct evaluation of the protein energy as a
function of its sequence.

Here, we use the data from Ref.~\onlinecite{zhou2005coarse} for the
so-called zinc-finger protein fold and closely follow exactly the same
computational procedures as employed in that study. The fitting is
done using a basis of approximately 76,000 clusters and energies of
60,000 AA sequences; a separate set of 4,000 AA sequence energies is
used to test the predictive power of the CE model. The very large size
of the problem presents a severe test to the conventional LOOCV-based
model building approach, requiring running times of several weeks on
parallel computers with user-supervised partial
optimization.~\cite{zhou2005coarse} We chose the highly efficient
split Bregman iteration\cite{goldstein2009split} for solving the basis
pursuit denoising problem in Eq.(\ref{eq:const_min_w_mu}), which
allows us to perform a full optimization in approximately 30 minutes
on a single 2.4 GHz Intel Xeon E5620
processor. Figure~\ref{fig:zinc-finger} shows that for the physically
important negative-energy configurations, we are able to achieve an
RMS predictive error of 2.1 kcal/mol with 3,100 model parameters,
significantly better than the RMS error of 2.7 kcal/mol with
approximately 6,000 parameters obtained using the LOOCV method in
Ref.~\onlinecite{zhou2005coarse}.  Since the predictive errors are
Gaussian-distributed with a mean of zero, the statistical uncertainty
in the predictive error due to the finite size of the prediction set
($>1000$ negative-energy structures) can be calculated using standard
statistical formulas for the $\chi^2$-distribution; they are
found to be less than 1\% of the calculated RMSE. These results show
that the computational efficiency, conceptual simplicity and physical
accuracy of the $\ell_1$-based minimization shows promise for future
applications in protein design.

\section{Conclusion}
In conclusion, compressive sensing can be straightforwardly adopted to
build physical models that are dominated by a relatively small number
of contributions drawn from a much larger underlying set of basis
functions.  Compressive sensing is applicable to any ``sparse''
basis-expansion problem, a broad class of problems in physics,
chemistry and materials science. Compressive sensing allows the
identification of relevant parameters from a large pool of candidates
using a small number of experiments or calculations---a real paradigm
shift from traditional techniques.  Furthermore, many other scientific
problems that do not appear to be a basis pursuit problem may be
recast as one, in which case CS could efficiently provide accurate and
robust solutions with relatively little user input.  With the huge
amount of experimental and computational data in physical sciences,
compressive sensing techniques represent a promising avenue for model
building on many fronts including structure maps, empirical potential
models, tight binding methods, and cluster expansions for
configurational energies, thermodynamics and kinetic Monte Carlo.

In the arena of cluster expansion, compressive sensing provides a
simple solution to two challenges: ``cluster selection'' and
``structure selection.'' Cluster selection is effectively solved
because compressive sensing can select clusters efficiently from a
very large set (thousands or tens of thousands). Essentially, it
allows the user to specify a cluster set so large that it encompasses
every physically-conceivable interaction. The second challenge,
structure selection, is overcome by the fact that compressive sensing
\emph{requires} that input structures simply be chosen randomly from
configuration space.

\hspace{1cm}

\section*{Acknowledgments}
We acknowledge several helpful suggestions from Stanley Osher
regarding the split Bregman method. G. L. W. H. and L. J. N. are
grateful for financial support from the NSF, DMR-0908753. F. Z. and
V. O. were supported as part of ``Molecularly Engineered Energy
Materials,'' an Energy Frontier Research Center funded by the
U.S. Department of Energy, Office of Science, Basic Energy Sciences
under Award Number DE-SC0001342 and used computing resources at the
National Energy Research Scientific Computing Center, which is
supported by the US DOE under Contract No. DE-AC02-05CH11231.


\begin{thebibliography}{60}%
\makeatletter
\providecommand \@ifxundefined [1]{%
 \@ifx{#1\undefined}
}%
\providecommand \@ifnum [1]{%
 \ifnum #1\expandafter \@firstoftwo
 \else \expandafter \@secondoftwo
 \fi
}%
\providecommand \@ifx [1]{%
 \ifx #1\expandafter \@firstoftwo
 \else \expandafter \@secondoftwo
 \fi
}%
\providecommand \natexlab [1]{#1}%
\providecommand \enquote  [1]{``#1''}%
\providecommand \bibnamefont  [1]{#1}%
\providecommand \bibfnamefont [1]{#1}%
\providecommand \citenamefont [1]{#1}%
\providecommand \href@noop [0]{\@secondoftwo}%
\providecommand \href [0]{\begingroup \@sanitize@url \@href}%
\providecommand \@href[1]{\@@startlink{#1}\@@href}%
\providecommand \@@href[1]{\endgroup#1\@@endlink}%
\providecommand \@sanitize@url [0]{\catcode `\\12\catcode `\$12\catcode
  `\&12\catcode `\#12\catcode `\^12\catcode `\_12\catcode `\%12\relax}%
\providecommand \@@startlink[1]{}%
\providecommand \@@endlink[0]{}%
\providecommand \url  [0]{\begingroup\@sanitize@url \@url }%
\providecommand \@url [1]{\endgroup\@href {#1}{\urlprefix }}%
\providecommand \urlprefix  [0]{URL }%
\providecommand \Eprint [0]{\href }%
\providecommand \doibase [0]{http://dx.doi.org/}%
\providecommand \selectlanguage [0]{\@gobble}%
\providecommand \bibinfo  [0]{\@secondoftwo}%
\providecommand \bibfield  [0]{\@secondoftwo}%
\providecommand \translation [1]{[#1]}%
\providecommand \BibitemOpen [0]{}%
\providecommand \bibitemStop [0]{}%
\providecommand \bibitemNoStop [0]{.\EOS\space}%
\providecommand \EOS [0]{\spacefactor3000\relax}%
\providecommand \BibitemShut  [1]{\csname bibitem#1\endcsname}%
\let\auto@bib@innerbib\@empty
\bibitem [{\citenamefont {Zunger}(1980)}]{Zunger1980}%
  \BibitemOpen
  \bibfield  {author} {\bibinfo {author} {\bibfnamefont {A.}~\bibnamefont
  {Zunger}},\ }\href@noop {} {\bibfield  {journal} {\bibinfo  {journal} {Phys.
  Rev. B}\ }\textbf {\bibinfo {volume} {22}},\ \bibinfo {pages} {5839}
  (\bibinfo {year} {1980})}\BibitemShut {NoStop}%
\bibitem [{\citenamefont {Villars}(1983)}]{Villars1983}%
  \BibitemOpen
  \bibfield  {author} {\bibinfo {author} {\bibfnamefont {P.}~\bibnamefont
  {Villars}},\ }\href@noop {} {\bibfield  {journal} {\bibinfo  {journal}
  {Journal of the Less Common Metals}\ }\textbf {\bibinfo {volume} {92}},\
  \bibinfo {pages} {215 } (\bibinfo {year} {1983})}\BibitemShut {NoStop}%
\bibitem [{\citenamefont {Pettifor}(1984)}]{Pettifor1984}%
  \BibitemOpen
  \bibfield  {author} {\bibinfo {author} {\bibfnamefont {D.}~\bibnamefont
  {Pettifor}},\ }\href@noop {} {\bibfield  {journal} {\bibinfo  {journal}
  {Solid State Communications}\ }\textbf {\bibinfo {volume} {51}},\ \bibinfo
  {pages} {31 } (\bibinfo {year} {1984})}\BibitemShut {NoStop}%
\bibitem [{\citenamefont {Pettifor}(1986)}]{Pettifor1986}%
  \BibitemOpen
  \bibfield  {author} {\bibinfo {author} {\bibfnamefont {D.}~\bibnamefont
  {Pettifor}},\ }\href@noop {} {\bibfield  {journal} {\bibinfo  {journal}
  {Journal of Physics C - Solid State Physics}\ }\textbf {\bibinfo {volume}
  {19}},\ \bibinfo {pages} {285} (\bibinfo {year} {1986})}\BibitemShut
  {NoStop}%
\bibitem [{\citenamefont {Miedema}\ \emph {et~al.}(1975)\citenamefont
  {Miedema}, \citenamefont {Boom},\ and\ \citenamefont
  {De~Boer}}]{Miedema1975}%
  \BibitemOpen
  \bibfield  {author} {\bibinfo {author} {\bibfnamefont {A.}~\bibnamefont
  {Miedema}}, \bibinfo {author} {\bibfnamefont {R.}~\bibnamefont {Boom}}, \
  and\ \bibinfo {author} {\bibfnamefont {F.~R.}\ \bibnamefont {De~Boer}},\
  }\href@noop {} {\bibfield  {journal} {\bibinfo  {journal} {Journal of the
  Less Common Metals}\ }\textbf {\bibinfo {volume} {41}},\ \bibinfo {pages}
  {283 } (\bibinfo {year} {1975})}\BibitemShut {NoStop}%
\bibitem [{\citenamefont {K\"ormann}\ \emph {et~al.}(2008)\citenamefont
  {K\"ormann}, \citenamefont {Dick}, \citenamefont {Grabowski}, \citenamefont
  {Hallstedt}, \citenamefont {Hickel},\ and\ \citenamefont
  {Neugebauer}}]{Kormann2008}%
  \BibitemOpen
  \bibfield  {author} {\bibinfo {author} {\bibfnamefont {F.}~\bibnamefont
  {K\"ormann}}, \bibinfo {author} {\bibfnamefont {A.}~\bibnamefont {Dick}},
  \bibinfo {author} {\bibfnamefont {B.}~\bibnamefont {Grabowski}}, \bibinfo
  {author} {\bibfnamefont {B.}~\bibnamefont {Hallstedt}}, \bibinfo {author}
  {\bibfnamefont {T.}~\bibnamefont {Hickel}}, \ and\ \bibinfo {author}
  {\bibfnamefont {J.}~\bibnamefont {Neugebauer}},\ }\href {\doibase
  10.1103/PhysRevB.78.033102} {\bibfield  {journal} {\bibinfo  {journal} {Phys.
  Rev. B}\ }\textbf {\bibinfo {volume} {78}},\ \bibinfo {pages} {033102}
  (\bibinfo {year} {2008})}\BibitemShut {NoStop}%
\bibitem [{\citenamefont {Sanchez}\ \emph {et~al.}(1984)\citenamefont
  {Sanchez}, \citenamefont {Ducastelle},\ and\ \citenamefont
  {Gratias}}]{sanchez1984generalized}%
  \BibitemOpen
  \bibfield  {author} {\bibinfo {author} {\bibfnamefont {J.}~\bibnamefont
  {Sanchez}}, \bibinfo {author} {\bibfnamefont {F.}~\bibnamefont {Ducastelle}},
  \ and\ \bibinfo {author} {\bibfnamefont {D.}~\bibnamefont {Gratias}},\
  }\href@noop {} {\bibfield  {journal} {\bibinfo  {journal} {Physica A:
  Statistical and Theoretical Physics}\ }\textbf {\bibinfo {volume} {128}},\
  \bibinfo {pages} {334} (\bibinfo {year} {1984})}\BibitemShut {NoStop}%
\bibitem [{\citenamefont {Fontaine}(1994)}]{fontaine1994cluster}%
  \BibitemOpen
  \bibfield  {author} {\bibinfo {author} {\bibfnamefont {D.}~\bibnamefont
  {Fontaine}},\ }\href@noop {} {\bibfield  {journal} {\bibinfo  {journal}
  {Solid State Physics}\ }\textbf {\bibinfo {volume} {47}},\ \bibinfo {pages}
  {33} (\bibinfo {year} {1994})}\BibitemShut {NoStop}%
\bibitem [{\citenamefont {Zunger}(1994)}]{Zunger1994}%
  \BibitemOpen
  \bibfield  {author} {\bibinfo {author} {\bibfnamefont {A.}~\bibnamefont
  {Zunger}},\ }\enquote {\bibinfo {title} {First-principles statistical
  mechanics of semiconductor alloys and intermetallic compounds},}\ \ (\bibinfo
   {publisher} {NATO Advanced Study Institute on Statics and Dynamics of Alloy
  Phase Transformations},\ \bibinfo {year} {1994})\ pp.\ \bibinfo {pages}
  {361--419}\BibitemShut {NoStop}%
\bibitem [{\citenamefont {Fischer}\ \emph {et~al.}(2006)\citenamefont
  {Fischer}, \citenamefont {Tibbetts}, \citenamefont {Morgan},\ and\
  \citenamefont {Ceder}}]{Fischer:2006ip}%
  \BibitemOpen
  \bibfield  {author} {\bibinfo {author} {\bibfnamefont {C.~C.}\ \bibnamefont
  {Fischer}}, \bibinfo {author} {\bibfnamefont {K.~J.}\ \bibnamefont
  {Tibbetts}}, \bibinfo {author} {\bibfnamefont {D.}~\bibnamefont {Morgan}}, \
  and\ \bibinfo {author} {\bibfnamefont {G.}~\bibnamefont {Ceder}},\
  }\href@noop {} {\bibfield  {journal} {\bibinfo  {journal} {Nature Materials}\
  }\textbf {\bibinfo {volume} {5}},\ \bibinfo {pages} {641} (\bibinfo {year}
  {2006})}\BibitemShut {NoStop}%
\bibitem [{\citenamefont {Sch\"on}\ \emph {et~al.}(2010)\citenamefont
  {Sch\"on}, \citenamefont {Doll},\ and\ \citenamefont
  {Jansen}}]{PSSB:PSSB200945246}%
  \BibitemOpen
  \bibfield  {author} {\bibinfo {author} {\bibfnamefont {J.~C.}\ \bibnamefont
  {Sch\"on}}, \bibinfo {author} {\bibfnamefont {K.}~\bibnamefont {Doll}}, \
  and\ \bibinfo {author} {\bibfnamefont {M.}~\bibnamefont {Jansen}},\ }\href
  {\doibase 10.1002/pssb.200945246} {\bibfield  {journal} {\bibinfo  {journal}
  {physica status solidi (b)}\ }\textbf {\bibinfo {volume} {247}},\ \bibinfo
  {pages} {23} (\bibinfo {year} {2010})}\BibitemShut {NoStop}%
\bibitem [{\citenamefont {Munter}\ \emph {et~al.}(2009)\citenamefont {Munter},
  \citenamefont {Landis}, \citenamefont {Abild-Pedersen}, \citenamefont
  {Jones}, \citenamefont {Wang},\ and\ \citenamefont
  {Bligaard}}]{1749-4699-2-1-015006}%
  \BibitemOpen
  \bibfield  {author} {\bibinfo {author} {\bibfnamefont {T.~R.}\ \bibnamefont
  {Munter}}, \bibinfo {author} {\bibfnamefont {D.~D.}\ \bibnamefont {Landis}},
  \bibinfo {author} {\bibfnamefont {F.}~\bibnamefont {Abild-Pedersen}},
  \bibinfo {author} {\bibfnamefont {G.}~\bibnamefont {Jones}}, \bibinfo
  {author} {\bibfnamefont {S.}~\bibnamefont {Wang}}, \ and\ \bibinfo {author}
  {\bibfnamefont {T.}~\bibnamefont {Bligaard}},\ }\href
  {http://stacks.iop.org/1749-4699/2/i=1/a=015006} {\bibfield  {journal}
  {\bibinfo  {journal} {Computational Science \& Discovery}\ }\textbf {\bibinfo
  {volume} {2}},\ \bibinfo {pages} {015006} (\bibinfo {year}
  {2009})}\BibitemShut {NoStop}%
\bibitem [{\citenamefont {Setyawan}\ \emph {et~al.}(2011)\citenamefont
  {Setyawan}, \citenamefont {Gaume}, \citenamefont {Lam}, \citenamefont
  {Feigelson},\ and\ \citenamefont {Curtarolo}}]{stefano_scintillator}%
  \BibitemOpen
  \bibfield  {author} {\bibinfo {author} {\bibfnamefont {W.}~\bibnamefont
  {Setyawan}}, \bibinfo {author} {\bibfnamefont {R.~M.}\ \bibnamefont {Gaume}},
  \bibinfo {author} {\bibfnamefont {S.}~\bibnamefont {Lam}}, \bibinfo {author}
  {\bibfnamefont {R.~S.}\ \bibnamefont {Feigelson}}, \ and\ \bibinfo {author}
  {\bibfnamefont {S.}~\bibnamefont {Curtarolo}},\ }\href {\doibase
  10.1021/co200012w} {\bibfield  {journal} {\bibinfo  {journal} {ACS
  Combinatorial Science}\ }\textbf {\bibinfo {volume} {13}},\ \bibinfo {pages}
  {382} (\bibinfo {year} {2011})},\ \Eprint
  {http://arxiv.org/abs/http://pubs.acs.org/doi/pdf/10.1021/co200012w}
  {http://pubs.acs.org/doi/pdf/10.1021/co200012w} \BibitemShut {NoStop}%
\bibitem [{\citenamefont {Woodley}\ and\ \citenamefont
  {Catlow}(2008)}]{Woodley:2008hha}%
  \BibitemOpen
  \bibfield  {author} {\bibinfo {author} {\bibfnamefont {S.~M.}\ \bibnamefont
  {Woodley}}\ and\ \bibinfo {author} {\bibfnamefont {R.}~\bibnamefont
  {Catlow}},\ }\href@noop {} {\bibfield  {journal} {\bibinfo  {journal} {Nature
  Materials}\ }\textbf {\bibinfo {volume} {7}},\ \bibinfo {pages} {937}
  (\bibinfo {year} {2008})}\BibitemShut {NoStop}%
\bibitem [{\citenamefont {Jansen}\ and\ \citenamefont
  {Popa}(2008)}]{jansen2008bayesian}%
  \BibitemOpen
  \bibfield  {author} {\bibinfo {author} {\bibfnamefont {A.~P.~J.}\
  \bibnamefont {Jansen}}\ and\ \bibinfo {author} {\bibfnamefont
  {C.}~\bibnamefont {Popa}},\ }\href {\doibase 10.1103/PhysRevB.78.085404}
  {\bibfield  {journal} {\bibinfo  {journal} {Phys.\ Rev.\ B}\ }\textbf
  {\bibinfo {volume} {78}},\ \bibinfo {pages} {085404} (\bibinfo {year}
  {2008})}\BibitemShut {NoStop}%
\bibitem [{\citenamefont {M\"uller}\ and\ \citenamefont
  {Ceder}(2009)}]{mueller2009bayesian}%
  \BibitemOpen
  \bibfield  {author} {\bibinfo {author} {\bibfnamefont {T.}~\bibnamefont
  {M\"uller}}\ and\ \bibinfo {author} {\bibfnamefont {G.}~\bibnamefont
  {Ceder}},\ }\href@noop {} {\bibfield  {journal} {\bibinfo  {journal} {Phys.\
  Rev.\ B}\ }\textbf {\bibinfo {volume} {80}},\ \bibinfo {pages} {024103}
  (\bibinfo {year} {2009})}\BibitemShut {NoStop}%
\bibitem [{\citenamefont {Cockayne}\ and\ \citenamefont {Van
  De~Walle}(2010)}]{cockayne2010building}%
  \BibitemOpen
  \bibfield  {author} {\bibinfo {author} {\bibfnamefont {E.}~\bibnamefont
  {Cockayne}}\ and\ \bibinfo {author} {\bibfnamefont {A.}~\bibnamefont {Van
  De~Walle}},\ }\href@noop {} {\bibfield  {journal} {\bibinfo  {journal}
  {Phys.\ Rev.\ B}\ }\textbf {\bibinfo {volume} {81}},\ \bibinfo {pages}
  {012104} (\bibinfo {year} {2010})}\BibitemShut {NoStop}%
\bibitem [{\citenamefont {Cand{\`e}s}\ and\ \citenamefont
  {Wakin}(2008)}]{candes2008introduction}%
  \BibitemOpen
  \bibfield  {author} {\bibinfo {author} {\bibfnamefont {E.}~\bibnamefont
  {Cand{\`e}s}}\ and\ \bibinfo {author} {\bibfnamefont {M.}~\bibnamefont
  {Wakin}},\ }\href@noop {} {\bibfield  {journal} {\bibinfo  {journal} {Signal
  Processing Magazine, IEEE}\ }\textbf {\bibinfo {volume} {25}},\ \bibinfo
  {pages} {21} (\bibinfo {year} {2008})}\BibitemShut {NoStop}%
\bibitem [{\citenamefont {AlQuraishi}\ and\ \citenamefont
  {McAdams}(2011)}]{alquraishi2011direct}%
  \BibitemOpen
  \bibfield  {author} {\bibinfo {author} {\bibfnamefont {M.}~\bibnamefont
  {AlQuraishi}}\ and\ \bibinfo {author} {\bibfnamefont {H.}~\bibnamefont
  {McAdams}},\ }\href@noop {} {\bibfield  {journal} {\bibinfo  {journal}
  {Proceedings of the National Academy of Sciences}\ }\textbf {\bibinfo
  {volume} {108}},\ \bibinfo {pages} {14819} (\bibinfo {year}
  {2011})}\BibitemShut {NoStop}%
\bibitem [{\citenamefont {Cand{\`e}s}\ \emph {et~al.}(2006)\citenamefont
  {Cand{\`e}s}, \citenamefont {Romberg},\ and\ \citenamefont
  {Tao}}]{candes2006robust}%
  \BibitemOpen
  \bibfield  {author} {\bibinfo {author} {\bibfnamefont {E.}~\bibnamefont
  {Cand{\`e}s}}, \bibinfo {author} {\bibfnamefont {J.}~\bibnamefont {Romberg}},
  \ and\ \bibinfo {author} {\bibfnamefont {T.}~\bibnamefont {Tao}},\
  }\href@noop {} {\bibfield  {journal} {\bibinfo  {journal} {Information
  Theory, IEEE Transactions on}\ }\textbf {\bibinfo {volume} {52}},\ \bibinfo
  {pages} {489} (\bibinfo {year} {2006})}\BibitemShut {NoStop}%
\bibitem [{\citenamefont {van~de Walle}\ \emph {et~al.}(2002)\citenamefont
  {van~de Walle}, \citenamefont {Asta},\ and\ \citenamefont
  {Ceder}}]{vandeWalle2002}%
  \BibitemOpen
  \bibfield  {author} {\bibinfo {author} {\bibfnamefont {A.}~\bibnamefont
  {van~de Walle}}, \bibinfo {author} {\bibfnamefont {M.}~\bibnamefont {Asta}},
  \ and\ \bibinfo {author} {\bibfnamefont {G.}~\bibnamefont {Ceder}},\ }\href
  {\doibase 10.1016/S0364-5916(02)80006-2} {\bibfield  {journal} {\bibinfo
  {journal} {Calphad}\ }\textbf {\bibinfo {volume} {26}},\ \bibinfo {pages}
  {539} (\bibinfo {year} {2002})}\BibitemShut {NoStop}%
\bibitem [{\citenamefont {van~de Walle}(2009)}]{vandeWalle2009}%
  \BibitemOpen
  \bibfield  {author} {\bibinfo {author} {\bibfnamefont {A.}~\bibnamefont
  {van~de Walle}},\ }\href {\doibase 10.1016/j.calphad.2008.12.005} {\bibfield
  {journal} {\bibinfo  {journal} {Calphad}\ }\textbf {\bibinfo {volume} {33}},\
  \bibinfo {pages} {266} (\bibinfo {year} {2009})}\BibitemShut {NoStop}%
\bibitem [{\citenamefont {Lerch}\ \emph {et~al.}(2009)\citenamefont {Lerch},
  \citenamefont {Wieckhorst}, \citenamefont {Hart}, \citenamefont {Forcade},\
  and\ \citenamefont {M\"uller}}]{lerch2009uncle}%
  \BibitemOpen
  \bibfield  {author} {\bibinfo {author} {\bibfnamefont {D.}~\bibnamefont
  {Lerch}}, \bibinfo {author} {\bibfnamefont {O.}~\bibnamefont {Wieckhorst}},
  \bibinfo {author} {\bibfnamefont {G.~L.~W.}\ \bibnamefont {Hart}}, \bibinfo
  {author} {\bibfnamefont {R.~W.}\ \bibnamefont {Forcade}}, \ and\ \bibinfo
  {author} {\bibfnamefont {S.}~\bibnamefont {M\"uller}},\ }\href@noop {}
  {\bibfield  {journal} {\bibinfo  {journal} {Modelling and Simulation in
  Materials Science and Engineering}\ }\textbf {\bibinfo {volume} {17}},\
  \bibinfo {pages} {055003} (\bibinfo {year} {2009})}\BibitemShut {NoStop}%
\bibitem [{\citenamefont {Zarkevich}\ and\ \citenamefont
  {Johnson}(2004)}]{Johnson2004}%
  \BibitemOpen
  \bibfield  {author} {\bibinfo {author} {\bibfnamefont {N.~A.}\ \bibnamefont
  {Zarkevich}}\ and\ \bibinfo {author} {\bibfnamefont {D.~D.}\ \bibnamefont
  {Johnson}},\ }\href {\doibase 10.1103/PhysRevLett.92.255702} {\bibfield
  {journal} {\bibinfo  {journal} {Phys. Rev. Lett.}\ }\textbf {\bibinfo
  {volume} {92}},\ \bibinfo {pages} {255702} (\bibinfo {year}
  {2004})}\BibitemShut {NoStop}%
\bibitem [{\citenamefont {Mueller}\ and\ \citenamefont
  {Ceder}(2010)}]{Ceder2010}%
  \BibitemOpen
  \bibfield  {author} {\bibinfo {author} {\bibfnamefont {T.}~\bibnamefont
  {Mueller}}\ and\ \bibinfo {author} {\bibfnamefont {G.}~\bibnamefont
  {Ceder}},\ }\href {\doibase 10.1103/PhysRevB.82.184107} {\bibfield  {journal}
  {\bibinfo  {journal} {Phys. Rev. B}\ }\textbf {\bibinfo {volume} {82}},\
  \bibinfo {pages} {184107} (\bibinfo {year} {2010})}\BibitemShut {NoStop}%
\bibitem [{\citenamefont {Connolly}\ and\ \citenamefont
  {Williams}(1983)}]{ConnollyWilliams}%
  \BibitemOpen
  \bibfield  {author} {\bibinfo {author} {\bibfnamefont {J.}~\bibnamefont
  {Connolly}}\ and\ \bibinfo {author} {\bibfnamefont {A.}~\bibnamefont
  {Williams}},\ }\href@noop {} {\bibfield  {journal} {\bibinfo  {journal}
  {Physical Review B}\ }\textbf {\bibinfo {volume} {27}},\ \bibinfo {pages}
  {5169} (\bibinfo {year} {1983})}\BibitemShut {NoStop}%
\bibitem [{\citenamefont {Blum}\ \emph {et~al.}(2005)\citenamefont {Blum},
  \citenamefont {Hart}, \citenamefont {Walorski},\ and\ \citenamefont
  {Zunger}}]{blum2005using}%
  \BibitemOpen
  \bibfield  {author} {\bibinfo {author} {\bibfnamefont {V.}~\bibnamefont
  {Blum}}, \bibinfo {author} {\bibfnamefont {G.~L.~W.}\ \bibnamefont {Hart}},
  \bibinfo {author} {\bibfnamefont {M.~J.}\ \bibnamefont {Walorski}}, \ and\
  \bibinfo {author} {\bibfnamefont {A.}~\bibnamefont {Zunger}},\ }\href@noop {}
  {\bibfield  {journal} {\bibinfo  {journal} {Phys.\ Rev.\ B}\ }\textbf
  {\bibinfo {volume} {72}},\ \bibinfo {pages} {165113} (\bibinfo {year}
  {2005})}\BibitemShut {NoStop}%
\bibitem [{\citenamefont {Laks}\ \emph {et~al.}(1992)\citenamefont {Laks},
  \citenamefont {Ferreira}, \citenamefont {Froyen},\ and\ \citenamefont
  {Zunger}}]{DBLaks}%
  \BibitemOpen
  \bibfield  {author} {\bibinfo {author} {\bibfnamefont {D.}~\bibnamefont
  {Laks}}, \bibinfo {author} {\bibfnamefont {L.}~\bibnamefont {Ferreira}},
  \bibinfo {author} {\bibfnamefont {S.}~\bibnamefont {Froyen}}, \ and\ \bibinfo
  {author} {\bibfnamefont {A.}~\bibnamefont {Zunger}},\ }\href@noop {}
  {\bibfield  {journal} {\bibinfo  {journal} {Physical Review B}\ }\textbf
  {\bibinfo {volume} {46}},\ \bibinfo {pages} {12587} (\bibinfo {year}
  {1992})}\BibitemShut {NoStop}%
\bibitem [{\citenamefont {Kurta}\ \emph {et~al.}(2010)\citenamefont {Kurta},
  \citenamefont {Bugaev},\ and\ \citenamefont {Ortiz}}]{Bugaev2010}%
  \BibitemOpen
  \bibfield  {author} {\bibinfo {author} {\bibfnamefont {R.~P.}\ \bibnamefont
  {Kurta}}, \bibinfo {author} {\bibfnamefont {V.~N.}\ \bibnamefont {Bugaev}}, \
  and\ \bibinfo {author} {\bibfnamefont {A.~D.}\ \bibnamefont {Ortiz}},\ }\href
  {\doibase 10.1103/PhysRevLett.104.085502} {\bibfield  {journal} {\bibinfo
  {journal} {Phys. Rev. Lett.}\ }\textbf {\bibinfo {volume} {104}},\ \bibinfo
  {pages} {085502} (\bibinfo {year} {2010})}\BibitemShut {NoStop}%
\bibitem [{\citenamefont {Thuinet}\ and\ \citenamefont
  {Besson}(2012)}]{Besson2012}%
  \BibitemOpen
  \bibfield  {author} {\bibinfo {author} {\bibfnamefont {L.}~\bibnamefont
  {Thuinet}}\ and\ \bibinfo {author} {\bibfnamefont {R.}~\bibnamefont
  {Besson}},\ }\href {\doibase 10.1063/1.4729426} {\bibfield  {journal}
  {\bibinfo  {journal} {Applied Physics Letters}\ }\textbf {\bibinfo {volume}
  {100}},\ \bibinfo {pages} {251902 } (\bibinfo {year} {2012})}\BibitemShut
  {NoStop}%
\bibitem [{\citenamefont {Shchyglo}\ \emph {et~al.}(2008)\citenamefont
  {Shchyglo}, \citenamefont {D{\'\i}az-Ortiz}, \citenamefont {Udyansky},
  \citenamefont {Bugaev}, \citenamefont {Reichert}, \citenamefont {Dosch},\
  and\ \citenamefont {Drautz}}]{DiazOrtiz2008}%
  \BibitemOpen
  \bibfield  {author} {\bibinfo {author} {\bibfnamefont {O.}~\bibnamefont
  {Shchyglo}}, \bibinfo {author} {\bibfnamefont {A.}~\bibnamefont
  {D{\'\i}az-Ortiz}}, \bibinfo {author} {\bibfnamefont {A.}~\bibnamefont
  {Udyansky}}, \bibinfo {author} {\bibfnamefont {V.~N.}\ \bibnamefont
  {Bugaev}}, \bibinfo {author} {\bibfnamefont {H.}~\bibnamefont {Reichert}},
  \bibinfo {author} {\bibfnamefont {H.}~\bibnamefont {Dosch}}, \ and\ \bibinfo
  {author} {\bibfnamefont {R.}~\bibnamefont {Drautz}},\ }\href
  {http://stacks.iop.org/0953-8984/20/i=4/a=045207} {\bibfield  {journal}
  {\bibinfo  {journal} {Journal of Physics: Condensed Matter}\ }\textbf
  {\bibinfo {volume} {20}},\ \bibinfo {pages} {045207} (\bibinfo {year}
  {2008})}\BibitemShut {NoStop}%
\bibitem [{\citenamefont {Mueller}\ and\ \citenamefont
  {Ceder}(2009)}]{Mueller2009}%
  \BibitemOpen
  \bibfield  {author} {\bibinfo {author} {\bibfnamefont {T.}~\bibnamefont
  {Mueller}}\ and\ \bibinfo {author} {\bibfnamefont {G.}~\bibnamefont
  {Ceder}},\ }\href@noop {} {\bibfield  {journal} {\bibinfo  {journal}
  {Physical Review B}\ }\textbf {\bibinfo {volume} {80}},\ \bibinfo {pages}
  {024103} (\bibinfo {year} {2009})}\BibitemShut {NoStop}%
\bibitem [{\citenamefont {Tibshirani}(1996)}]{Tibshirani1996}%
  \BibitemOpen
  \bibfield  {author} {\bibinfo {author} {\bibfnamefont {R.}~\bibnamefont
  {Tibshirani}},\ }\href@noop {} {\bibfield  {journal} {\bibinfo  {journal} {J.
  Roy. Stat. Soc. Ser. B}\ }\textbf {\bibinfo {volume} {58}},\ \bibinfo {pages}
  {267} (\bibinfo {year} {1996})}\BibitemShut {NoStop}%
\bibitem [{\citenamefont {Chen}\ \emph {et~al.}(1998)\citenamefont {Chen},
  \citenamefont {Donoho},\ and\ \citenamefont {Saunders}}]{Chen1998}%
  \BibitemOpen
  \bibfield  {author} {\bibinfo {author} {\bibfnamefont {S.}~\bibnamefont
  {Chen}}, \bibinfo {author} {\bibfnamefont {D.}~\bibnamefont {Donoho}}, \ and\
  \bibinfo {author} {\bibfnamefont {M.}~\bibnamefont {Saunders}},\ }\href@noop
  {} {\bibfield  {journal} {\bibinfo  {journal} {SIAM J. Sci. Comput.}\
  }\textbf {\bibinfo {volume} {20}},\ \bibinfo {pages} {33} (\bibinfo {year}
  {1998})}\BibitemShut {NoStop}%
\bibitem [{\citenamefont {Hale}\ \emph {et~al.}(2007)\citenamefont {Hale},
  \citenamefont {Yin},\ and\ \citenamefont {Zhang}}]{HaleYinZhang2007}%
  \BibitemOpen
  \bibfield  {author} {\bibinfo {author} {\bibfnamefont {E.}~\bibnamefont
  {Hale}}, \bibinfo {author} {\bibfnamefont {W.}~\bibnamefont {Yin}}, \ and\
  \bibinfo {author} {\bibfnamefont {Y.}~\bibnamefont {Zhang}},\ }\href@noop {}
  {\bibfield  {journal} {\bibinfo  {journal} {CAAM TR07-07, Rice University}\ }
  (\bibinfo {year} {2007})}\BibitemShut {NoStop}%
\bibitem [{\citenamefont {Yin}\ \emph {et~al.}(2008)\citenamefont {Yin},
  \citenamefont {Osher}, \citenamefont {Goldfarb},\ and\ \citenamefont
  {Darbon}}]{yin2008bregman}%
  \BibitemOpen
  \bibfield  {author} {\bibinfo {author} {\bibfnamefont {W.}~\bibnamefont
  {Yin}}, \bibinfo {author} {\bibfnamefont {S.}~\bibnamefont {Osher}}, \bibinfo
  {author} {\bibfnamefont {D.}~\bibnamefont {Goldfarb}}, \ and\ \bibinfo
  {author} {\bibfnamefont {J.}~\bibnamefont {Darbon}},\ }\href@noop {}
  {\bibfield  {journal} {\bibinfo  {journal} {SIAM Journal on Imaging
  Sciences}\ }\textbf {\bibinfo {volume} {1}},\ \bibinfo {pages} {143}
  (\bibinfo {year} {2008})}\BibitemShut {NoStop}%
\bibitem [{\citenamefont {Boyd}\ and\ \citenamefont
  {Vandenberghe}(2004)}]{Boyd}%
  \BibitemOpen
  \bibfield  {author} {\bibinfo {author} {\bibfnamefont {S.}~\bibnamefont
  {Boyd}}\ and\ \bibinfo {author} {\bibfnamefont {L.}~\bibnamefont
  {Vandenberghe}},\ }\href@noop {} {\emph {\bibinfo {title} {{Convex
  Optimization}}}}\ (\bibinfo  {publisher} {Cambridge University Press},\
  \bibinfo {year} {2004})\BibitemShut {NoStop}%
\bibitem [{\citenamefont {Goldstein}\ and\ \citenamefont
  {Osher}(2009)}]{goldstein2009split}%
  \BibitemOpen
  \bibfield  {author} {\bibinfo {author} {\bibfnamefont {T.}~\bibnamefont
  {Goldstein}}\ and\ \bibinfo {author} {\bibfnamefont {S.}~\bibnamefont
  {Osher}},\ }\href@noop {} {\bibfield  {journal} {\bibinfo  {journal} {SIAM
  Journal on Imaging Sciences}\ }\textbf {\bibinfo {volume} {2}},\ \bibinfo
  {pages} {323} (\bibinfo {year} {2009})}\BibitemShut {NoStop}%
\bibitem [{\citenamefont {Donoho}\ and\ \citenamefont
  {Huo}(2001)}]{donoho2001}%
  \BibitemOpen
  \bibfield  {author} {\bibinfo {author} {\bibfnamefont {D.}~\bibnamefont
  {Donoho}}\ and\ \bibinfo {author} {\bibfnamefont {X.}~\bibnamefont {Huo}},\
  }\href@noop {} {\bibfield  {journal} {\bibinfo  {journal} {IEEE Transactions
  on Information Theory}\ }\textbf {\bibinfo {volume} {47}},\ \bibinfo {pages}
  {2845} (\bibinfo {year} {2001})}\BibitemShut {NoStop}%
\bibitem [{\citenamefont {Candes}\ and\ \citenamefont
  {Romberg}(2007)}]{Candes}%
  \BibitemOpen
  \bibfield  {author} {\bibinfo {author} {\bibfnamefont {E.}~\bibnamefont
  {Candes}}\ and\ \bibinfo {author} {\bibfnamefont {J.}~\bibnamefont
  {Romberg}},\ }\href@noop {} {\bibfield  {journal} {\bibinfo  {journal}
  {Inverse problems}\ }\textbf {\bibinfo {volume} {23}},\ \bibinfo {pages}
  {969} (\bibinfo {year} {2007})}\BibitemShut {NoStop}%
\bibitem [{\citenamefont {Candes}\ \emph {et~al.}(2006)\citenamefont {Candes},
  \citenamefont {Romberg},\ and\ \citenamefont {Tao}}]{candes2006stable}%
  \BibitemOpen
  \bibfield  {author} {\bibinfo {author} {\bibfnamefont {E.}~\bibnamefont
  {Candes}}, \bibinfo {author} {\bibfnamefont {J.}~\bibnamefont {Romberg}}, \
  and\ \bibinfo {author} {\bibfnamefont {T.}~\bibnamefont {Tao}},\ }\href@noop
  {} {\bibfield  {journal} {\bibinfo  {journal} {Communications on pure and
  applied mathematics}\ }\textbf {\bibinfo {volume} {59}},\ \bibinfo {pages}
  {1207} (\bibinfo {year} {2006})}\BibitemShut {NoStop}%
\bibitem [{\citenamefont {Hart}\ and\ \citenamefont
  {Forcade}(2008)}]{hart2008algorithm}%
  \BibitemOpen
  \bibfield  {author} {\bibinfo {author} {\bibfnamefont {G.~L.~W.}\
  \bibnamefont {Hart}}\ and\ \bibinfo {author} {\bibfnamefont {R.~W.}\
  \bibnamefont {Forcade}},\ }\href@noop {} {\bibfield  {journal} {\bibinfo
  {journal} {Phys.\ Rev.\ B}\ }\textbf {\bibinfo {volume} {77}},\ \bibinfo
  {pages} {224115} (\bibinfo {year} {2008})}\BibitemShut {NoStop}%
\bibitem [{\citenamefont {Hart}\ and\ \citenamefont
  {Forcade}(2009)}]{hart2009generating}%
  \BibitemOpen
  \bibfield  {author} {\bibinfo {author} {\bibfnamefont {G.~L.~W.}\
  \bibnamefont {Hart}}\ and\ \bibinfo {author} {\bibfnamefont {R.~W.}\
  \bibnamefont {Forcade}},\ }\href@noop {} {\bibfield  {journal} {\bibinfo
  {journal} {Phys.\ Rev.\ B}\ }\textbf {\bibinfo {volume} {80}},\ \bibinfo
  {pages} {014120} (\bibinfo {year} {2009})}\BibitemShut {NoStop}%
\bibitem [{\citenamefont {Van~de Walle}\ \emph {et~al.}(2002)\citenamefont
  {Van~de Walle}, \citenamefont {Asta},\ and\ \citenamefont {Ceder}}]{atat}%
  \BibitemOpen
  \bibfield  {author} {\bibinfo {author} {\bibfnamefont {A.}~\bibnamefont
  {Van~de Walle}}, \bibinfo {author} {\bibfnamefont {M.}~\bibnamefont {Asta}},
  \ and\ \bibinfo {author} {\bibfnamefont {G.}~\bibnamefont {Ceder}},\
  }\href@noop {} {\bibfield  {journal} {\bibinfo  {journal} {Calphad}\ }\textbf
  {\bibinfo {volume} {26}},\ \bibinfo {pages} {539} (\bibinfo {year}
  {2002})}\BibitemShut {NoStop}%
\bibitem [{\citenamefont {Seko}\ \emph {et~al.}(2009)\citenamefont {Seko},
  \citenamefont {Koyama},\ and\ \citenamefont {Tanaka}}]{seko2009cluster}%
  \BibitemOpen
  \bibfield  {author} {\bibinfo {author} {\bibfnamefont {A.}~\bibnamefont
  {Seko}}, \bibinfo {author} {\bibfnamefont {Y.}~\bibnamefont {Koyama}}, \ and\
  \bibinfo {author} {\bibfnamefont {I.}~\bibnamefont {Tanaka}},\ }\href@noop {}
  {\bibfield  {journal} {\bibinfo  {journal} {Phys.\ Rev.\ B}\ }\textbf
  {\bibinfo {volume} {80}},\ \bibinfo {pages} {165122} (\bibinfo {year}
  {2009})}\BibitemShut {NoStop}%
\bibitem [{ddj()}]{ddj_fraction_factorial_design}%
  \BibitemOpen
  \href@noop {} {}\bibinfo {note} {Teck Tan and Duane Johnson recently
  developed an as-yet-unpublished method for structure selection that applies
  fractional factorial design to the structure selection problem (private
  communication)}\BibitemShut {NoStop}%
\bibitem [{\citenamefont {Johnstone}(2001)}]{Johnstone2001}%
  \BibitemOpen
  \bibfield  {author} {\bibinfo {author} {\bibfnamefont {I.}~\bibnamefont
  {Johnstone}},\ }\href@noop {} {\bibfield  {journal} {\bibinfo  {journal} {The
  Annals of statistics}\ }\textbf {\bibinfo {volume} {29}},\ \bibinfo {pages}
  {295} (\bibinfo {year} {2001})}\BibitemShut {NoStop}%
\bibitem [{Note1()}]{Note1}%
  \BibitemOpen
  \bibinfo {note} {This applies only to random numerical errors in the DFT
  formation energies and excludes systematic errors, such as those due to the
  approximate nature of the exchange-correlation functionals.}\BibitemShut
  {Stop}%
\bibitem [{Note2()}]{Note2}%
  \BibitemOpen
  \bibinfo {note} {In our estimation, numerical errors in the calculated DFT
  formation energies are only a few meV/atom for the case of Ag-Pt compounds
  considered in Sec.~\ref {sec:AgPt}.}\BibitemShut {Stop}%
\bibitem [{\citenamefont {Durussel}\ and\ \citenamefont
  {Feschotte}(1996)}]{Durussel1996}%
  \BibitemOpen
  \bibfield  {author} {\bibinfo {author} {\bibfnamefont {P.}~\bibnamefont
  {Durussel}}\ and\ \bibinfo {author} {\bibfnamefont {P.}~\bibnamefont
  {Feschotte}},\ }\href@noop {} {\bibfield  {journal} {\bibinfo  {journal} {J.\
  Alloys Compound.}\ }\textbf {\bibinfo {volume} {239}},\ \bibinfo {pages}
  {226} (\bibinfo {year} {1996})}\BibitemShut {NoStop}%
\bibitem [{Note3()}]{Note3}%
  \BibitemOpen
  \bibinfo {note} {Such a large number of structures was only chosen to test
  the performance of different CE methods and is several times larger than
  typical training set sizes used in state-of-the-art CE methods.}\BibitemShut
  {Stop}%
\bibitem [{\citenamefont {Kresse}\ and\ \citenamefont
  {Joubert}(1999)}]{Kresse:1999wc}%
  \BibitemOpen
  \bibfield  {author} {\bibinfo {author} {\bibfnamefont {G.}~\bibnamefont
  {Kresse}}\ and\ \bibinfo {author} {\bibfnamefont {D.}~\bibnamefont
  {Joubert}},\ }\href@noop {} {\bibfield  {journal} {\bibinfo  {journal}
  {Phys.\ Rev.\ B}\ }\textbf {\bibinfo {volume} {59}},\ \bibinfo {pages} {1758}
  (\bibinfo {year} {1999})}\BibitemShut {NoStop}%
\bibitem [{\citenamefont {Kresse}\ and\ \citenamefont
  {Furthm\"uller}(1996)}]{kresse1996efficiency}%
  \BibitemOpen
  \bibfield  {author} {\bibinfo {author} {\bibfnamefont {G.}~\bibnamefont
  {Kresse}}\ and\ \bibinfo {author} {\bibfnamefont {J.}~\bibnamefont
  {Furthm\"uller}},\ }\href@noop {} {\bibfield  {journal} {\bibinfo  {journal}
  {Comp.\ Mat.\ Sci.}\ }\textbf {\bibinfo {volume} {6}},\ \bibinfo {pages} {15}
  (\bibinfo {year} {1996})}\BibitemShut {NoStop}%
\bibitem [{\citenamefont {Bl\"ochl}(1994)}]{Blochl:1994dx}%
  \BibitemOpen
  \bibfield  {author} {\bibinfo {author} {\bibfnamefont {P.~E.}\ \bibnamefont
  {Bl\"ochl}},\ }\href@noop {} {\bibfield  {journal} {\bibinfo  {journal}
  {Phys.\ Rev.\ B}\ }\textbf {\bibinfo {volume} {50}},\ \bibinfo {pages}
  {17953} (\bibinfo {year} {1994})}\BibitemShut {NoStop}%
\bibitem [{\citenamefont {Perdew}\ \emph {et~al.}(1996)\citenamefont {Perdew},
  \citenamefont {Burke},\ and\ \citenamefont {Ernzerhof}}]{PBExc}%
  \BibitemOpen
  \bibfield  {author} {\bibinfo {author} {\bibfnamefont {J.~P.}\ \bibnamefont
  {Perdew}}, \bibinfo {author} {\bibfnamefont {K.}~\bibnamefont {Burke}}, \
  and\ \bibinfo {author} {\bibfnamefont {M.}~\bibnamefont {Ernzerhof}},\ }\href
  {\doibase 10.1103/PhysRevLett.77.3865} {\bibfield  {journal} {\bibinfo
  {journal} {Phys. Rev. Lett.}\ }\textbf {\bibinfo {volume} {77}},\ \bibinfo
  {pages} {3865} (\bibinfo {year} {1996})}\BibitemShut {NoStop}%
\bibitem [{\citenamefont {Froyen}(1989)}]{froyen1989brillouin}%
  \BibitemOpen
  \bibfield  {author} {\bibinfo {author} {\bibfnamefont {S.}~\bibnamefont
  {Froyen}},\ }\href@noop {} {\bibfield  {journal} {\bibinfo  {journal} {Phys.\
  Rev.\ B}\ }\textbf {\bibinfo {volume} {39}},\ \bibinfo {pages} {3168}
  (\bibinfo {year} {1989})}\BibitemShut {NoStop}%
\bibitem [{\citenamefont {Nguyen}\ and\ \citenamefont
  {Stehl{\'e}}(2009)}]{Minkowski}%
  \BibitemOpen
  \bibfield  {author} {\bibinfo {author} {\bibfnamefont {P.~Q.}\ \bibnamefont
  {Nguyen}}\ and\ \bibinfo {author} {\bibfnamefont {D.}~\bibnamefont
  {Stehl{\'e}}},\ }\href {\doibase 10.1145/1597036.1597050} {\bibfield
  {journal} {\bibinfo  {journal} {ACM Trans. Algorithms}\ }\textbf {\bibinfo
  {volume} {5}},\ \bibinfo {pages} {1} (\bibinfo {year} {2009})}\BibitemShut
  {NoStop}%
\bibitem [{\citenamefont {Nelson}\ \emph {et~al.}(2012)\citenamefont {Nelson},
  \citenamefont {Hart},\ and\ \citenamefont {Curtarolo}}]{nelson2012ground}%
  \BibitemOpen
  \bibfield  {author} {\bibinfo {author} {\bibfnamefont {L.}~\bibnamefont
  {Nelson}}, \bibinfo {author} {\bibfnamefont {G.}~\bibnamefont {Hart}}, \ and\
  \bibinfo {author} {\bibfnamefont {S.}~\bibnamefont {Curtarolo}},\ }\href@noop
  {} {\bibfield  {journal} {\bibinfo  {journal} {Phys.\ Rev.\ B}\ }\textbf
  {\bibinfo {volume} {85}},\ \bibinfo {pages} {054203} (\bibinfo {year}
  {2012})}\BibitemShut {NoStop}%
\bibitem [{\citenamefont {Zhou}\ \emph {et~al.}(2005)\citenamefont {Zhou},
  \citenamefont {Grigoryan}, \citenamefont {Lustig}, \citenamefont {Keating},
  \citenamefont {Ceder},\ and\ \citenamefont {Morgan}}]{zhou2005coarse}%
  \BibitemOpen
  \bibfield  {author} {\bibinfo {author} {\bibfnamefont {F.}~\bibnamefont
  {Zhou}}, \bibinfo {author} {\bibfnamefont {G.}~\bibnamefont {Grigoryan}},
  \bibinfo {author} {\bibfnamefont {S.}~\bibnamefont {Lustig}}, \bibinfo
  {author} {\bibfnamefont {A.}~\bibnamefont {Keating}}, \bibinfo {author}
  {\bibfnamefont {G.}~\bibnamefont {Ceder}}, \ and\ \bibinfo {author}
  {\bibfnamefont {D.}~\bibnamefont {Morgan}},\ }\href@noop {} {\bibfield
  {journal} {\bibinfo  {journal} {Physical review letters}\ }\textbf {\bibinfo
  {volume} {95}},\ \bibinfo {pages} {148103} (\bibinfo {year}
  {2005})}\BibitemShut {NoStop}%
\bibitem [{\citenamefont {Grigoryan}\ \emph {et~al.}(2009)\citenamefont
  {Grigoryan}, \citenamefont {Reinke},\ and\ \citenamefont
  {Keating}}]{grigoryan2009design}%
  \BibitemOpen
  \bibfield  {author} {\bibinfo {author} {\bibfnamefont {G.}~\bibnamefont
  {Grigoryan}}, \bibinfo {author} {\bibfnamefont {A.}~\bibnamefont {Reinke}}, \
  and\ \bibinfo {author} {\bibfnamefont {A.}~\bibnamefont {Keating}},\
  }\href@noop {} {\bibfield  {journal} {\bibinfo  {journal} {Nature}\ }\textbf
  {\bibinfo {volume} {458}},\ \bibinfo {pages} {859} (\bibinfo {year}
  {2009})}\BibitemShut {NoStop}%
\end{thebibliography}
%

\end{document}